\documentclass[a4paper,11pt]{article}
\pdfoutput=1
\usepackage{jheppub}
\usepackage[english]{babel}
\usepackage{amsthm,amsfonts}
\allowdisplaybreaks
\usepackage[mathscr]{euscript}
\usepackage{enumitem}
\usepackage{tikz}
\usetikzlibrary{arrows}
\usetikzlibrary{calc}
\usetikzlibrary{decorations.pathreplacing}
\usetikzlibrary{matrix}
\tikzset{>=latex}
\usepackage{dsfont}
\usepackage{bbm}
\DeclareMathAlphabet{\mathdsl}{U}{bbm}{m}{sl}
\usepackage{ytableau}
\usepackage{cancel}

\usepackage{cite}

\DeclareMathOperator{\asym}{asym}
\newtheorem{theorem}{Theorem}

\newcommand{\dd}{\mathrm{d}}
\newcommand{\Pb}{\overline{P}{}}
\newcommand{\WZW}{\mathrm{WZW}}
\newcommand{\s}{$\sigma$}
\newcommand{\PS}{Pol\'a\v{c}ek-Siegel}
\newcommand{\fh}{\widehat{f}}
\newcommand{\Eh}{\widehat{E}}
\newcommand{\Ah}{\widehat{A}}
\newcommand{\Bh}{\widehat{B}}
\newcommand{\Ch}{\widehat{C}}
\newcommand{\Dh}{\widehat{D}}
\newcommand{\Fh}{\widehat{F}}

\newcommand{\Ih}{\widehat{I}}
\newcommand{\Jh}{\widehat{J}}
\newcommand{\ah}{\widehat{a}}
\newcommand{\bh}{\widehat{b}}
\newcommand{\ch}{\widehat{c}}
\renewcommand{\dh}{\widehat{d}}
\newcommand{\db}{\overline{d}}
\renewcommand{\fh}{\widehat{f}}
\newcommand{\ih}{\widehat{i}}
\newcommand{\jh}{\widehat{j}}
\newcommand{\vh}{\widehat{v}}
\newcommand{\lambdah}{\widehat{\lambda}}
\newcommand{\vt}{\widetilde{v}}
\newcommand{\DD}{\mathds{D}}
\newcommand{\LL}{\mathcal{L}}
\newcommand{\HH}{\mathcal{H}}
\newcommand{\JJ}{\mathcal{J}}
\newcommand{\EE}{\mathcal{E}}
\newcommand{\RR}{\mathcal{R}}
\newcommand{\GG}{\mathcal{G}}
\newcommand{\nablaB}{\overline{\nabla}}
\newcommand{\GammaB}{\overline{\Gamma}}
\newcommand{\FF}{\mathcal{F}}
\newcommand{\Dc}{\mathcal{D}}
\newcommand{\Rb}{\overline{R}}
\newcommand{\Et}{\widetilde{E}}
\newcommand{\Kt}{\widetilde{K}}
\newcommand{\Tt}{\widetilde{T}}
\newcommand{\phib}{\overline{\phi}}
\newcommand{\dt}{\widetilde{d}}
\newcommand{\alphad}{\dot\alpha}
\newcommand{\betad}{\dot\beta}
\newcommand{\gammad}{\dot\gamma}
\newcommand{\Vh}{\widehat{V}}
\newcommand{\Eb}{\overline{E}}

\newcommand{\VV}{\mathcal{V}}
\newcommand{\rr}{\mathbf{r}}
\newcommand{\Gt}{\widetilde{G}}
\newcommand{\et}{\widetilde{e}}
\newcommand{\rmm}{\mathrm{m}}
\newcommand{\Mt}{\widetilde{M}}
\newcommand{\Mb}{\overline{M}}
\newcommand{\Bb}{\overline{B}}
\newcommand{\Hb}{\overline{H}}
\renewcommand{\vt}{\widetilde{v}}
\newcommand{\vth}{\widehat{\vt}}
\newcommand{\vb}{\overline{v}}

\newcommand{\yt}{\widetilde{y}}
\newcommand{\zt}{\widetilde{z}}
\newcommand{\tF}{t_{\mathrm{F}}}

\title{\boldmath Consistent Truncations and Dualities}
\preprint{MI-HET-788}

\author[a]{Daniel Butter,}
\author[a,b]{Falk Hassler,}
\author[a]{Christopher N. Pope,}
\author[a]{and Haoyu Zhang}

\emailAdd{dbutter@tamu.edu}
\emailAdd{falk.hassler@uwr.edu.pl}
\emailAdd{pope@physics.tamu.edu}
\emailAdd{zhanghaoyu@tamu.edu}

\affiliation[a]{George P. \& Cynthia Woods Mitchell Institute for Fundamental Physics and Astronomy,\\ Texas A\&M University, College Station, TX 77843, USA}
\affiliation[b]{University of Wrocław, Faculty of Physics and Astronomy, Maksa Borna 9, 50-204 Wrocław, Poland}

\abstract{Recent progress in generalised geometry and extended field theories suggests a deep connection between consistent truncations and dualities, which is not immediately obvious. A prime example is generalised Scherk-Schwarz reductions in double field theory, which have been shown to be in one-to-one correspondence with Poisson-Lie T-duality. Here we demonstrate that this relation is only the tip of the iceberg. Currently, the most general known classes of T-dualities (excluding mirror symmetry) are based on dressing cosets. But as we discuss, they can be further extended to the even larger class of generalised cosets. We prove that the latter give rise to consistent truncations for which the ansatz can be constructed systematically. Hence, we pave the way for many new examples of T-dualities and consistent truncations. The arising structures result in covariant tensors with more than two derivatives and we argue how they might be key to understand generalised T-dualities and consistent truncations beyond the leading two derivative level.}

\begin{document}

\maketitle

\section{Introduction}

The notions of consistent truncations and dualities are at first sight 
seemingly unrelated.
The former singles out degrees of freedom in a physical theory which 
decouple from the rest. They are important because it is often 
easier to analyse a system, i.e. solve its equations of motion, if the 
number of degrees of freedom is reduced. Therefore, consistent truncations 
provide a crucial tool for finding solutions in (super)gravity, 
with a wide range of applications. More precisely, this idea can be 
summarised by the commuting diagram 
\begin{equation}\label{diag:contrunc}
  \tikz{
    \node (fullaction) {full action $S$};
    \node[at={(fullaction.east)},anchor=center,xshift=20em] (redaction) {truncated action $S_{\mathrm{red}}$};
    \node[at={(fullaction.south)},anchor=center,yshift=-3em] (fieldeq) {field equations $\delta S = 0$};
    \node[at={(redaction.south)},anchor=center,yshift=-3em] (redfieldeq) {truncated field equations $\delta S_{\mathrm{red}} = 0$.};
    \draw[->] (fullaction.east) -- (redaction.west);
    \draw[->] (fullaction.south) -- (fieldeq.north);
    \draw[->] (redaction.south) -- (redfieldeq.north);
    \draw[->] (fieldeq.east) -- (redfieldeq.west);
    \draw[->] ($(fieldeq.south) + (0,-0.5)$) -- ($(redfieldeq.south) + (0,-0.5)$) node[midway,above] {truncation};
    \draw[<-] ($(fieldeq.south) + (0,-0.6)$) -- ($(redfieldeq.south) + (0,-0.6)$) node[midway,below] {uplift};
  }
\end{equation}
The truncation is said to be  
consistent if the two pathways to the truncated equations of motion 
yield the same result. Otherwise, the chosen truncation ansatz does not 
single out decoupled degrees of freedom. One should note that it is in 
general very difficult to find consistent truncations, because the 
standard Kaluza-Klein ansatz with massless gauge fields 
is in general inconsistent \cite{Duff:1984hn}. For a long time, only 
few exceptions have been known, including sphere reductions 
\cite{Cvetic:2000dm} and reductions on group manifolds \cite{Scherk:1979zr}.

The second central concept for this paper is dualities. They are ubiquitous 
in physics, and find applications from models in condensed matter to 
high energy physics. The basic idea is that two seemingly very different 
models ultimately still share the same (quantum/classical) dynamics. Here, 
we are particularly interested in target-space dualities (T-dualities) 
of two-dimensional \s-models. They can be studied from two major 
perspectives: the worldsheet and the target space. The former is the surface 
on which the \s-model is defined as a field theory. In the classical limit, 
T-duality acts as a canonical transformation relating at least two different 
\s-models on different target spaces. Alternatively, one can consider the 
low-energy, effective target-space action that captures the dynamics of the 
strings described by the \s-model. Here, T-duality maps existing solutions 
of the field equations to new solutions. In this context it plays a role as 
a solution generating technique. It is known that a small subset of all 
T-dualities, namely abelian T-dualities, are preserved under quantum 
corrections on the worldsheet and the corresponding higher-derivative 
corrections in the target space effective action. For the remaining, 
generalised, T-dualities, their fate under quantum corrections is not 
yet known. There are some preliminary results 
\cite{Hassler:2020tvz,Borsato:2020wwk,Codina:2020yma,Hassler:2020wnp} 
that suggest that they might also cover higher derivative corrections. 
Here, we are mostly concerned with the leading two-derivative effective 
action, and such higher-derivative corrections will just touch the discussion 
tangentially.

Historically, one distinguishes between non-abelian T-duality 
\cite{delaOssa:1992vci}, Poisson-Lie T-duality 
\cite{Klimcik:1995ux,Klimcik:1995dy}, 
which might be supplemented by a Wess-Zumino-Witten term 
\cite{Klimcik:2001vg}, and dressing cosets \cite{Klimcik:1996np}. 
Our results will apply to all of these, and we shall call them
{\it generalised T-dualities}. Note that we will not consider 
mirror symmetries, which relate different Calabi-Yau manifolds; they are 
related to abelian T-duality by the SYZ conjecture \cite{Strominger:1996it}.
For completeness, let us quickly explain the two central
mathematical objects mentioned here. A Poisson-Lie group is a Lie group $G$
equipped with a Poisson bracket that satisfies
\begin{equation}
  \{f_1, f_2\}(g g') = \{f_1 \circ L_g, f_2 \circ L_g \}(g') + 
    \{ f_1 \circ R_{g'}, f_2 \circ R_{g'} \}(g)
\end{equation}
where $L_h(g) = h g$ denotes the left-multiplication on $G$ and
$R_h (g) = g h$ the right-multiplication, respectively. The classical Lie
group $\rightarrow$ Lie algebra correspondence was extended by Drinfeld to
Poisson-Lie group $\rightarrow$ Lie bialgebra, which contains in
addition to the Lie algebra Lie($G$) also a dual Lie algebra Lie($\Gt$)
corresponding to dual Lie group $\Gt$. Together, these two Lie groups
form a double Lie group $\DD$ with the Lie algebra
Lie($\DD$)=Lie($G$)$\oplus$Lie($\Gt$). Poisson-Lie groups are called dual,
if they share the same $\DD$. For example, they can be related by
exchanging $G$ and $\Gt$. Poisson-Lie T-duality got its
name because it identifies \s-models whose target spaces are dual Poisson-Lie
groups. In addition to the right- and left-action of $G$ on $G$, Poisson-Lie groups admit a so-called dressing action \cite{10.4310/jdg/1214444324}. It can be most easily seen on the level of the doubled Lie group $\DD$ where any element of the subgroup $F \subset \DD$ generates the dressing transformation $g\rightarrow g'$ on $G$ with
\begin{equation}
  f g = g' \widetilde{h} \,, \qquad g,\,g'\in G\,, \quad 
  \widetilde{h}\in\Gt\,, \quad \text{and} \quad
  f \in F\,.
\end{equation}
Here the group multiplication is the multiplication on $\DD$. The set of orbits of this action is called the dressing coset \cite{Klimcik:1996np}. In this paper, we will even work with a more general notion than dressing cosets, which are called generalised cosets \cite{Demulder:2019vvh}. They drop the requirement that the doubled Lie group $\DD$ has to originate from a Poisson-Lie group and can be understood as the lift of the concept of a coset in differential geometry to generalised geometry. At the end, the following dependencies arise:
\begin{equation}
  \begin{array}{ccccccc}
    \text{ abelian } & \subset & \text{ non-abelian } & \subset & \text{ Poisson-Lie } & \subset & \text{ WZW-Poisson }\\
    & & & & 
    \raisebox{0.6em}{\rotatebox{270}{$\subset$}}
    & &
    \raisebox{0.6em}{\rotatebox{270}{$\subset$}}\\
    & & & & \text{ dressing coset } & \subset & \text{ \underline{generalised coset}\,. }
  \end{array}
\end{equation}

At first glance these two concepts, consistent truncations and generalised 
T-duality, seem unrelated. However, in recent years it has become
evident that a deeper understanding of each of these concepts 
can be achieved using similar tools, most prominently (exceptional) 
generalised geometry \cite{Coimbra:2011nw,Coimbra:2012af} and 
double/exceptional field theories. In particular the latter were 
initially developed \cite{Siegel:1993th,0904.4664,Hohm:2010pp} 
with abelian T-duality (and later, its extension to U-duality) in mind; 
before its utility for understanding 
consistent truncations was appreciated 
\cite{Lee:2014mla,Hohm:2014qga,Cassani:2019vcl}. While double field theory 
is in principle able to describe also non-geometric setups, we 
shall use it in its most conservative form, with the standard 
solution $\widetilde{\partial}^i = 0$ of the section condition, thus 
rendering it equivalent to generalised geometry. Later, it was shown that 
generalised T-dualities can also be conveniently studied in double 
field theory \cite{Hassler:2017yza,Demulder:2018lmj,Sakatani:2019jgu}. 
Moreover, they provided the first systematic construction of the generalised 
frame fields that underlie generalised Scherk-Schwarz reductions
\cite{Dabholkar:2002sy,Aldazabal:2011nj,Geissbuhler:2011mx}, and which still
dominate the landscape of consistent truncations. Motivated by this
connection and examples like \cite{Itsios:2012dc}, the goal of this paper is to explore further the relation between
generalised T-dualities and consistent truncations. It is already known that
all generalised T-dualities except for dressing cosets give rise to consistent 
generalised Scherk-Schwarz reductions. Therefore, we shall take a closer 
look at generalised cosets, and show how they can be used to construct a very 
rich class of consistent truncations that go beyond generalised Scherk-Schwarz 
reductions.

 Our starting point will be the most general known class of 
consistent truncations 
\cite{Cassani:2019vcl} that arise from generalised geometry. 
In section~\ref{sec:ggconsistent}, we review the underlying construction. 
At the end of this section, we then show how the already mentioned 
generalised Scherk-Schwarz reductions are connected to Poisson-Lie T-duality. 
At this point, we already see that dualities can be understood naturally 
in the context of consistent truncations, by relating different 
truncation ans\"atze for a higher-dimensional theory that give rise to 
the same truncated theory:
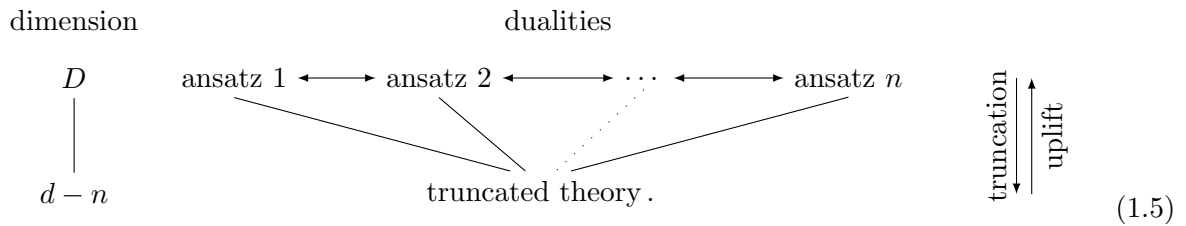
\begin{equation}\label{eqn:dualitiesdiagram}
  \begin{tikzpicture}
    \node[name=ansatz1] {ansatz 1};
    \node[at={(ansatz1)},xshift=7em,name=ansatz2] {ansatz 2};
    \node[at={(ansatz2)},xshift=7em,name=dots] {\dots};
    \node[at={(dots)},xshift=7em,name=ansatzn] {ansatz $n$};
    \draw[<->] (ansatz1.east) -- (ansatz2.west);
    \draw[<->] (ansatz2.east) -- (dots.west) node[midway,yshift=2em] {dualities};
    \draw[<->] (dots.east) -- (ansatzn.west);
    \path (ansatz1.east) -- (ansatzn.west) node[midway,yshift=-4em,name=trunc] {truncated theory\,.};
    \draw (ansatz1.south) -- ($(trunc.north)-(0.4,0)$);
    \draw (ansatz2.south) -- ($(trunc.north)-(0.2,0)$);
    \draw[loosely dotted] (dots.south) -- ($(trunc.north)+(0.2,0)$);
    \draw (ansatzn.south) -- ($(trunc.north)+(0.4,0)$);
    \node [at={($(trunc.west)-(4.5,0)$)},name=dminusn] {$d-n$};
    \node [at={(dminusn)},yshift=4em,name=D] {$D$};
    \node [at={(D)},yshift=2em] {dimension};
    \draw (dminusn.north) -- (D.south);
    \draw[->] ($(dminusn)+(12.6,0)$) -- ($(D)+(12.6,0)$) node[midway,rotate=90,below] {uplift};
    \draw[<-] ($(dminusn)+(12.4,0)$) -- ($(D)+(12.4,0)$) node[midway,rotate=90,above] {truncation};
  \end{tikzpicture}
\end{equation}
\bigskip

Inspired by this finding, we are led to pose the question:
\begin{center}
  Is there a one-to-one correspondence between generalised T-dualities 
and consistent truncations in generalised geometry?
\end{center}
To answer it, section~\ref{sec:mega-vielbein} introduces a new construction 
for the truncation ans\"atze depicted in \eqref{eqn:dualitiesdiagram}. 
It employs a higher-dimensional, auxiliary space to geometrise the 
generalised structure group that underlies each consistent truncation. 
A similar technique was used by Pol\'a\v{c}ek and Siegel 
\cite{Polacek:2013nla} some time ago to find a natural construction of
the generalised Riemann tensor in double field theory. It provides
a structured construction of covariant torsion and curvature tensors which, 
as we shall show, are of central interest in consistent truncations also. 
With this tool at hand, we establish in section~\ref{sec:truncgenRicci}
that all spaces that admit 
generalised T-dualities give rise to consistent truncations. 
Establishing the converse, namely that any consistent truncation can be 
constructed from a dressing coset, is more involved. On the one hand, 
we know that consistent truncations can be constructed on Sasaki-Einstein 
spaces \cite{Cassani:2019vcl} that are clearly not double cosets. However, 
this is not, of itself, a problem, because as the illustration 
\eqref{eqn:dualitiesdiagram} shows, there can exist different ans\"atze 
that result in the same truncation. To check if at least one of them 
originates from a dressing coset we have to find solutions for the Jacobi 
identity of the underlying Lie algebra in which certain components of the 
structure coefficients are fixed while others remain free. We set up the 
computation in section~\ref{sec:Jac}, but it is hard to find solutions 
because the Jacobi identities lead to many coupled quadratic equations. 
This problem gets easier the more components of the structure coefficients 
are fixed, because they do not then appear as unknowns in the quadratic 
equations. Fortunately, this is exactly what happens when consistent 
truncations are considered in connection with higher-derivative corrections. 
We explore this idea in section~\ref{sec:higherderiv}, and interpret it as 
a hint that there might indeed be a one-to-one connection between 
consistent truncations and generalised T-dualities. We plan to return to
the problem of constructing a complete proof in future work. 
Finally, section~\ref{sec:vielbeins} is concerned with the explicit 
construction of truncation an\"atze for generalised cosets, the basis for 
any generalised T-duality. Here, we extend the results of 
previous work \cite{Demulder:2019vvh}, by presenting a completely 
systematic construction. This culminates with the 
observation~\ref{th:MAIN} that on any dressing coset 
$H \backslash \DD / F$ one can construct a consistent truncation with 
generalised structure group $F$.

\section{Generalised geometry and consistent truncations}\label{sec:ggconsistent}
\subsection{(Super)gravity and generalised geometry}\label{sec:gengeo}

We are interested in consistent truncations of (super)gravity, such as 
arises as the low-energy effective action in string theory. For the 
sake of simplicity, we only consider the NS/NS sector, which is governed 
by the action
\begin{equation}\label{eqn:SNSNS}
  S = \int \dd^D x\, \sqrt{g} e^{-2\phi} \left( R + 4 \partial_i \phi \partial^i \phi - \frac1{12} H_{ijk} H^{ijk} \right)\,.
\end{equation}
We use the convention here that $g$ is the determinant of the metric 
$g_{ij}$, and $R$ the corresponding curvature scalar. Besides the metric, 
there is also the dilaton $\phi$ and the $B$-field $B_{ij}$ on the 
$D$-dimensional spacetime $M_D$ (also referred to as the target space). 
The action does not incorporate $B_{ij}$ directly, but instead its field 
strength $H_{ijk} = 3 \partial_{[i} B_{jk]}$. Conformal invariance of the 
string at the quantum level fixes $D=10$ for superstrings and 
$D=26$ for their bosonic counterpart. \eqref{eqn:SNSNS} possesses two 
local symmetries, namely diffeomorphisms that account for coordinate 
changes and $B$-field gauge transformations 
$B_{ij} \rightarrow B_{ij} + 2 \partial_{[i} \xi_{j]}$. 

The infinitesimal versions of these two symmetries 
can be written in a unified form in terms of the generalised Lie derivative
\begin{equation}\label{eqn:genLie}
  \LL_U V^I = U^J \partial_J V^I - (\partial_J U^I - \partial^I U_J) V^J\,.
\end{equation}
$U^I = \begin{pmatrix} u^i \quad& u_i\end{pmatrix}$, 
$U_I = \begin{pmatrix} u_i \quad& u^i \end{pmatrix}$ and $V^I$ 
are generalised vectors on the generalised tangent space $T M \oplus T^* M$. 
While the original tangent space $T M$ is always extended in this setup, 
the manifold $M_D$ might be extended (double/exceptional field theories) or 
not (generalised geometry). The two alternatives are related by the section 
condition, which singles out the non-constant, physical, directions on $M_D$. 
We always solve the section condition in the trivial way, 
namely $\partial_I = \begin{pmatrix} \partial_i & 0 \end{pmatrix}$. 
Thus we do not need to distinguish between the two different approaches, 
and we just use the doubling of the partial derivative index as a convenient 
book-keeping device. The form of \eqref{eqn:genLie} is fixed by the 
requirement that the natural pairing between a vector and a one-form,
\begin{equation}\label{eqn:etametric}
  U^I V^J \eta_{IJ} =  u^i v_i + u_i v^i\,, \quad \text{with} 
\quad \eta_{IJ} = \begin{pmatrix} 0 & \delta_i^j \\ \delta_j^i & 0 
\end{pmatrix}\,,
\end{equation}
is preserved. The invariance of this pairing introduces an 
O($D$,$D$) structure. This structure allows the metric and 
$B$-field to be captured in terms of one unified object, 
the generalised metric
\begin{equation}\label{eqn:genmetric}
  \HH_{IJ} = \begin{pmatrix}
    g_{ij} - B_{ik} g^{kl} B_{lj} &\qquad -B_{ik} g^{kj} \\
    g^{ik} B_{kj} & \qquad g^{ij}
  \end{pmatrix}\,.
\end{equation}
It has two defining properties: 1) it is symmetric and 2) it is an 
element of O($D$,$D$);\ $\HH_{IK} \eta^{KL} \HH_{LJ} = \eta_{IJ}$, 
where $\eta^{IJ}$ is the inverse of $\eta_{IJ}$. Moreover, its 
parameterisation in terms of $g_{ij}$ and $B_{ij}$ is chosen such that 
the infinitesimal action of diffeomorphisms and $B$-field gauge 
transformation is mediated by the generalised Lie derivative. 

The generalised metric introduces more structure than might be immediately 
obvious from the parameterisation \eqref{eqn:genmetric}. To reveal it, we 
re-express $\HH_{IJ}$ in terms of the generalised frame field $E^A{}_I$, with the defining properties
\begin{align}
  \eta_{IJ} &= E^A{}_I \eta_{AB} E^B{}_J\,, & \eta_{AB} &= \begin{pmatrix} 0 & \delta_a^b \\ \delta_b^a & 0 \end{pmatrix} \quad \text{and}\\\label{eqn:HHcanonical}
  \HH_{IJ}  &= E^A{}_I \HH_{AB} E^B{}_J\,,  & \HH_{AB}  &= \begin{pmatrix} \eta_{ab} & 0 \\ 0 & \eta^{ab} \end{pmatrix} \,,
\end{align}
where $\eta_{ab}$ is either the Lorentzian or Euclidean metric and 
$\eta^{ab}$ its inverse. These two relations still do not fix the 
generalised frame completely. One can perform coordinate dependent double 
Lorentz transformations, $E_A{}^I \rightarrow \Lambda_A{}^B E_B{}^I$, 
without changing the resulting $\eta_{IJ}$ and $\HH_{IJ}$, assuming 
that $\Lambda_A{}^C \eta_{CD} \Lambda_B{}^D = 
\Lambda_A{}^C \HH_{CD} \Lambda_B{}^D = 0$. They furnish the double Lorentz 
group $H_D$=O($D$-1,1)$\times$O(1,$D$-1) for Lorentzian or 
$H_D$=O($D$)$\times$O($D$) for Euclidean spacetimes.

We can use these ingredients to rewrite the action \eqref{eqn:SNSNS}. 
There are in fact different ways to do this. For our purpose the so-called flux formulation is most suitable 
\cite{Hohm:2010xe,Geissbuhler:2011mx,Geissbuhler:2013uka}. To make the 
invariance under generalised diffeomorphisms (generated by the generalised 
Lie derivative) of the action manifest, the flux formulation introduces two 
generalised fluxes, namely
\begin{equation}\label{eqn:defgenfluxes}
  \LL_{E_A} E_B = F_{AB}{}^C E_C  \quad \text{and} \quad \LL_{E_A} e^{-2 d} = - F_A e^{-2 d} \,.
\end{equation}
Taking into account the definition of the generalised Lie derivative 
\eqref{eqn:genLie}, the first equation leads to
\begin{equation}\label{eqn:FABC}
  F_{ABC} = 3 E_{[A}{}^I \partial_I E_B{}^J E_{C]J}\,,
\end{equation}
while the second relation requires more explanation because we encounter 
a new quantity, $d$, the generalised dilaton. It is defined by
\begin{equation}
  e^{-2 d} = \sqrt{g} e^{- 2 \phi} \quad \text{or} \quad
  d = \phi - \frac14 \log g\,.
\end{equation}
Importantly, $e^{-2 d}$ does not transform as a scalar under 
the generalised Lie derivative, but rather, as a scalar density, resulting in
\begin{equation}
  \LL_U e^{- 2 d} = \partial_I ( U^I e^{-2 d} )\,.
\end{equation}
Therefore, the second generalised flux $F_A$ is given by
\begin{equation}\label{eqn:FA}
  F_A = 2 E_A{}^I \partial_I d - \partial_I E_A{}^I\,.
\end{equation}

The two fluxes $F_{ABC}$ and $F_A$
repackage the information contained in the metric, dilaton and $B$-field. 
Hence, the action \eqref{eqn:SNSNS} can be alternatively written in terms 
of them as \cite{Geissbuhler:2011mx,Geissbuhler:2013uka}
\begin{equation}\label{eqn:SDFT}
  S = \int \dd^D x \, e^{-2 d} \RR
\end{equation}
with the generalised Ricci scalar being given by
\begin{equation}\label{eqn:genRicciScalar}
  \RR = P^{AB} P^{CD} \left( \Pb^{EF} + \frac13 P^{EF} \right) F_{ACE} F_{BDF} + 2 P^{AB}( 2 D_A F_B - F_A F_B ) \,.
\end{equation}
We encounter two new objects here: First, the flat derivative
\begin{equation}
  D_A = E_A{}^I \partial_I
\end{equation}
and second, the projectors
\begin{equation}
  P^{AB} = \frac12 ( \eta^{AB} + \HH^{AB} ) \quad \text{and} \quad
  \Pb^{AB} = \frac12 ( \eta^{AB} - \HH^{AB} )\,.
\end{equation}
They are called projectors because of their properties
\begin{equation}
  P_A{}^C P_C{}^B = P_A{}^B \,, \quad
  \Pb_A{}^C \Pb_C{}^B = \Pb_A{}^B \,, \quad \text{and} \quad
  P_A{}^C \Pb_C{}^B = \Pb_A{}^C P_C{}^B = 0 \,.
\end{equation}
At this point, we just have a rewriting of \eqref{eqn:SNSNS} in terms of 
quantities that appear ``naturally'' in generalised geometry or 
double field theory.

\subsection{Systematics of consistent truncations}

Next, we explain how this form of the action helps in identifying 
consistent truncations. The answer is given by the following theorem,
which was established in \cite{Cassani:2019vcl}:
\begin{theorem}\label{th:CT}
Let $M_D$ be a $D$-dimensional manifold with a generalised $F$-structure 
defining a set of invariant tensors ${f^{(j)}}$ with $F \subset H_D$ and only constant, singlet intrinsic torsion. Then there is a consistent truncation of the action \eqref{eqn:SNSNS} on $M_D$ defined by expanding all bosonic fields in terms of invariant tensors.
\end{theorem}
\noindent A key observation to understand this theorem is that 
invariant tensors are covariantly constant with respect to an appropriate 
O($D$,$D$) covariant derivative $\nabla_A$, i.e. $\nabla_A f^{(j)} = 0$ for 
all $f^{(j)}$. This derivative acts as a selector for degrees of freedom 
that are retained in the truncation. An important feature of this  
derivative is that since it obeys the Leibniz rule, the product of any two 
invariant tensors is again covariantly constant, and thus will be 
part of the truncation. Usually, the derivative $\nabla_I$ is not the 
generalised Levi-Civita connection $\nablaB_I$ from which the generalised 
curvature scalar in the action \eqref{eqn:SDFT} is derived. Fortunately
though, the two are related. To see how, we have to look closer at 
covariant derivatives in generalised geometry/double field theory.

Different generalised covariant derivatives differ by their 
generalised torsion. Similarly to the situation in standard geometry, 
the torsion is defined by comparing the generalised Lie derivative written 
with 
partial or covariant derivatives as in \cite{Coimbra:2011nw}:
\begin{equation}\label{eqn:defgentorsion}
  \left( \LL^{\nabla}_U - \LL^{\partial}_U \right) V^I = U^J V^K T_{JK}{}^I \,.
\end{equation}
To compute this quantity directly, we need an explicit definition of the 
covariant derivative:
\begin{equation}\label{eqn:defnabla}
  \nabla_I E_A{}^J = \partial_I E_A{}^J - \Omega_{IA}{}^B E_B{}^J + 
\Gamma_{IK}{}^J E_A{}^K \,,
\end{equation}
involving the generalised spin connection $\Omega_{IA}{}^B E_B{}^J$ and 
the generalised affine connection $\Gamma_{IJ}{}^K$. The two are related by 
the vielbein postulate $\nabla_I E_A{}^J = 0$, leading to
\begin{equation}\label{eqn:GammafromOmega}
  \Gamma_{IJK} = \partial_I E^A{}_J E_{AK} + \Omega_{IJK} \,.
\end{equation}
With these definitions, the generalised torsion evaluates to
\begin{equation}\label{eqn:gentorsionTIJK}
  T_{IJK} = 3 \Gamma_{[IJK]} \,.
\end{equation}

Additionally, the generalised covariant derivative should at least 
satisfy two more constraints \cite{Coimbra:2011nw}:
\begin{enumerate}
  \item Compatibility with the $\eta$-metric, $\nabla_I \eta_{JK} = 0$, implying 
    \begin{equation}
      \Gamma_{IJK} = -\Gamma_{IKJ}\,.
    \end{equation}
  \item Compatibility with the generalised metric, $\nabla_I \HH_{JK} =0$, implying
    \begin{equation}
      P_{(J}{}^M \Pb_{K)}{}^N \Gamma_{IMN} = \frac12 P_{(J}{}^M \Pb_{K)}{}^N \partial_I \HH_{MN}\,.
    \end{equation}
  \item\label{item:IBP} (optional) Compatibility with integration by parts $\int \dd^D x \, e^{-2 d} \nabla_I V^I = 0$, implying
    \begin{equation}
      \GammaB^J{}_{JI} = 2 \partial_I d\,.
    \end{equation}
\end{enumerate}
Note that for the generalised Levi-Civita connection, point \ref{item:IBP} applies, but it does not have to hold for $\nabla_I$. In the definition of the generalised torsion \eqref{eqn:defgentorsion}, the generalised Lie derivative acts on a vector. We obtain a similar result for higher rank generalised tensors, where $T_{IJ}{}^K$ acts on each index individually. But there are also densities, like the generalised dilaton $d$. Hence, we introduce an additional torsion for them, too,
\begin{equation}
  \left( \LL^{\nabla}_U - \LL^{\nablaB}_U \right) e^{-2 d} = U^I T_I e^{-2 d} \,,
\end{equation}
which gives rise to
\begin{equation}\label{eqn:gentorsionTI}
  T_I = 2 \partial_I d - \Gamma^J{}_{JI}\,.
\end{equation}

As in conventional geometry, the generalised Levi-Civita connection, $\nablaB_I$, is defined to have vanishing torsion. Thus, we can deduce the relation between the generalised Christoffel symbols $\GammaB_{IJK}$ of $\nablaB_I$ and $\Gamma_{IJK}$ of $\nabla_I$,
\begin{equation}\label{eqn:defSigma}
  \begin{aligned}
    \Sigma_{IJK} &= \GammaB_{IJK} - \Gamma_{IJK} \\ 
                 &= -\left( \frac13 P P P + \Pb P P + P \Pb \Pb + \frac13 \Pb \Pb \Pb  \right){}_{IJK}{}^{LMN}
                 \left( T_{LMN} + \frac6{D-1} \eta_{L[M}T_{N]} \right) \,.
  \end{aligned}
\end{equation}
Note that $\Sigma_{IJK}$ is not fully fixed by all constraints above. There 
are some undetermined components \cite{Coimbra:2011nw}, which we have set 
to zero here. However, it is known that these components do not contribute 
to the two-derivative effective action \eqref{eqn:SNSNS} or its field 
equations. Therefore, it is safe to neglect them.

After this digression, we recognise that the action of the generalised Levi-Civita connection on the invariant tensors has to be of the form
\begin{equation}
  \nablaB_I f^{(j)} = \Sigma_I \cdot f^{(j)} \,.
\end{equation}
According to theorem~\ref{th:CT}, the torsions $T_{IJK}$ and 
$T_I$ of $\nabla$, called the intrinsic torsion, have to be covariantly 
constant. From this observation and \eqref{eqn:defSigma}, it is obvious that
\begin{equation}
  \nabla_I \Sigma_{JKL} = 0
\end{equation}
holds. Therefore, any tensor expression which is a function $u$ of $f^{(j)}$ and their $\nablaB$-derivatives satisfies
\begin{equation}
  \nabla_I u( f^{(j)}, \nablaB ) = 0
\end{equation}
automatically. Still, this does not guarantee that the resulting tensor is 
contained in the set ${f^{(j)}}$ that forms the truncation. To overcome this 
problem, we note that all elements of this set are invariant under the 
action of the group $F$. Any function $u( f^{(j)}, \nablaB )$ will share 
this property as long as the intrinsic torsion is a singlet (i.e.~it 
is invariant) under the $F$ action. In this case $\Sigma_{IJK}$ is invariant 
because, due to $F \subset H_D$, the projectors in its definition are 
automatically invariant, too.

Let us summarise the ingredients of the construction implied by 
theorem~\ref{th:CT} and fix some notation for later. For a 
consistent truncation we need:
\begin{enumerate}
  \item The set of \underline{all} invariant tensors $\{ f^{(j)} \}$ which satisfy
    \begin{equation}\label{eqn:definvartensors}
      \nabla_I f^{(j)} = 0 \quad \text{and} \quad \nabla_\alpha f^{(j)} = 0 \,.
    \end{equation}
    Here $\nabla_\alpha$ denotes the infinitesimal action by generators $t_\alpha \in$ Lie($F$) of the structure group $F$. The reason why we use this particular notation, will become clear in the next section.
  \item Constant, singlet intrinsic torsion that implies
    \begin{equation}\label{eqn:defsinglettorsion}
      \nabla_I T_{JKL} =  0\,,\quad \nabla_I T_J = 0  \qquad \text{and} \qquad 
      \nabla_\alpha T_{IJK} = 0\,, \quad \nabla_\alpha T_I = 0 \,.
    \end{equation}
\end{enumerate}

\subsection{The truncated theory}\label{sec:trunctheory}

At this point, we can say more about the truncated theory. In general, the 
consistent truncation still has an infinite number of degrees of freedom. 
To accommodate them, the $D$-dimensional target space is split into two 
parts
\begin{equation}
  M_D = M_{D-n} \times M_{n}\,,
\end{equation}
comprising an external manifold $M_{D-n}$ of dimension $D-n$ where no 
truncation is performed, and the internal manifold $M_n$ that hosts the 
invariant tensors discussed above. Accordingly, we split the coordinates, 
namely $x^\mu$ on $M_{D-n}$ and $y^i$ on $M_n$. On the external space, we do 
not require generalised geometry because no truncation takes place here. 
Therefore, we do not use doubled indices on $M_{D-n}$. Only on the internal 
space $M_n$, we have the O($n$,$n$)-invariant pairing $\eta_{IJ}$, a 
generalised metric $\HH_{IJ}$, and all the other objects introduced above. 
In this setup the metric, $B$-field and dilaton on the full space split 
into five contributions:
\begin{itemize}
  \item external metric $g_{\mu\nu}( x )$\,,
  \item external $B$-field $B_{\mu\nu}( x )$\,,
  \item dilaton $\phi(x, y)$\,,
  \item gauge connection $A_\mu{}^I(x, y)$\,, and
  \item scalar field $\HH_{IJ}(x, y)$\,.
\end{itemize}
All shall be understood as fields in the truncated theory with their $y$-dependence totally fixed by the truncation ansatz.

The action \eqref{eqn:SNSNS} can be rewritten in terms of these fields by 
using a Kaluza-Klein ansatz. This rewriting is cumbersome, but can be found 
in different papers, e.g. 
\cite{Aldazabal:2011nj,Geissbuhler:2011mx,Hohm:2013nja}. Here, we start from 
the result of \cite{Hohm:2013nja}, because it is fully general and does not 
commit to any specific ansatz on the internal manifold yet:
\begin{equation}\label{eqn:SNSNSsplit}
  \begin{aligned}
    S = \int \dd^{(D-n)} x\, \dd^n y \sqrt{g} e^{-2 \phi} \Bigl( & \Rb + \Dc_\mu \phi \Dc^\mu \phi - \frac1{12} \HH_{\mu\nu\rho} \HH^{\mu\nu\rho} \\
    & + \frac18 \Dc_\mu \HH_{IJ} \Dc^\mu \HH^{IJ} - \frac14 \HH_{IJ} \FF_{\mu\nu}{}^{I} \FF^{\mu\nu J} - V \Bigr)\,.
  \end{aligned}
\end{equation}
In the following, we will refine this action by using all the insights from 
the previous subsection. But before doing so, let us discuss the new objects 
that appear in \eqref{eqn:SNSNSsplit}. $\Rb$ is the curvature scalar for the 
external metric. Furthermore, the truncated theory is a gauge theory 
coupled to (super)gravity, normally referred to as a gauged (super)gravity. 
It comes with the gauge covariant derivative
\begin{equation}
  \Dc_\mu = \partial_\mu - \LL_{A_\mu}
\end{equation}
that incorporates the connection one-form $A_\mu{}^I$. Like in 
Yang-Mills theory, one has to add a kinetic term containing the corresponding 
2-form field strength tensor
\begin{equation}
  \FF_{\mu\nu}{}^I = 2 \partial_{[\mu} A_{\nu]}{}^I - [ A_\mu, A_\nu ]_{\mathrm{C}}^I
\end{equation}
to the action. Instead of a Lie bracket, it employs the Courant bracket
\begin{equation}
  [ U, V ]_{\mathrm{C}} = \frac12 (\LL_U V - \LL_V U)\,.
\end{equation}
This is because we have not yet performed the truncation. Later, we will 
see that this bracket turns into a Lie bracket after the truncation, 
as is required for a gauge theory. Moreover, the 3-form $H_{ijk}$ in the 
original action has to be complemented by a Chern-Simons term giving 
rise to
\begin{equation}
  \HH_{\mu\nu\rho} = 3 \partial_{[\mu} B_{\nu\rho]} + 3 A_{[\mu}{}^I \partial_{\nu} A_{\rho]I} - A_{[\mu| I |} [ A_{\nu}, A_{\rho]} ]_{\mathrm{C}}^I\,.
\end{equation}
Finally, we have the scalar potential
\begin{equation}
  V = - \RR\,.
\end{equation}
It is expressed in terms of the generalised Ricci scalar 
\eqref{eqn:genRicciScalar} \textit{on the internal manifold $M_n$}. 
For completeness, we note that gauge transformations of the connection 
1-form $A_\mu{}^I$ are mediated by
\begin{equation}
  \delta_\Lambda A_\mu{}^I = \partial_\mu \Lambda^I + \LL_\Lambda A_\mu{}^I \,,
\end{equation}
For all the covariant fields, the generalised Lie derivative mediates 
gauge transformation, namely
\begin{equation}
  \delta_\Lambda = \LL_\Lambda \,.
\end{equation}

At this stage, there are still an infinite number of scalars $\HH_{IJ}$ 
and vectors $A_\mu{}^I$. But, as suggested above, we should expand them in 
terms of tensors that are invariant under the structure group in order 
to obtain a consistent truncation. Let us start with the generalised metric
\begin{equation}
  \HH_{IJ}(x, y) = \Et^A{}_I(y) h_{AB}(x) \Et^B{}_J(y)\,,
\end{equation}
and adopt the notation that quantities which depend only on $y$ are 
decorated with a tilde. In contrast to section~\ref{sec:gengeo}, the 
generalised metric $h_{AB}$ here is not restricted to the diagonal form
given in \eqref{eqn:HHcanonical}. Rather, it is an element of the 
coset O($n$,$n$)/$H_n$ and it thereby captures the scalar moduli space of 
the truncated theory. However, this choice has to be refined because it is 
not automatically invariant under the action of the structure group $F$. 
Thus, we restrict $h_{AB}$ to the coset
\begin{equation}
  h_{AB}(x) \in \frac{C_{\mathrm{O}(n,n)}(F)}{C_{H_n}(F)}
\end{equation}
where $C_G(H)$ denotes all element of the Lie group $G$ that commute with all element of $H$. Similarly, the vectors of the truncated theory are formed from all $n_{\mathrm{v}}$ $F$-invariant vectors $\Kt_{\alphad}{}^I (y)$, with $\alphad = 1, \dots, n_{\mathrm{v}}$,
\begin{equation}
  A_\mu{}^I(x, y) = A_\mu{}^{\alphad} (x) \Kt_{\alphad}{}^I (y) \,.
\end{equation}
Thanks to the analysis in the last section, it is not hard to see how these vectors act on other invariant tensors through the generalised Lie derivative.\footnote{The trick here is to write the generalised Lie derivative in terms of the covariant derivative $\nabla_A$ and keep in mind that it annihilates all $F$-invariant tensors on the internal manifold. Thus, the only non-vanishing contribution comes from the torsion tensor.} In particular, we find
\begin{align}
  \LL_{\Kt_{\alphad}} \Kt_{\betad}{}^I &= (\Tt_{\alphad})_J{}^I \Kt_{\betad}{}^J \\
  \LL_{\Kt_{\betad}} \HH_{IJ} &= 2 (\Tt_{\alphad})_{(I}{}^K \HH_{J)K}
\end{align}
with
\begin{equation}
  (\Tt_{\alphad})_I{}^J = - \Kt_{\alphad}{}^K \Tt_{KI}{}^J\,.
\end{equation}
Here, $\Tt_{IJK}$ is the generalised torsion from \eqref{eqn:gentorsionTIJK}. We only decorated it with a tilde to emphasis that it depends only on the internal coordinates $y^i$. One can interpret $(\Tt_{\alphad})_I{}^J$ as $2n \times 2n$-matrices. They are elements of the Lie algebra Lie[$C_{\mathrm{O}(n,n)}(F)$] and generate the gauge group $G\subset C_{\mathrm{O}(n,n)}(F)$ \cite{Cassani:2019vcl}. The corresponding Lie algebra has the structure \textit{constants}
\begin{equation}
  [ \Tt_{\alphad}, \Tt_{\betad} ] = f_{\alphad\betad}{}^{\dot\gamma} \Tt_{\dot\gamma}\,.
\end{equation}
This result is quite remarkable, because it tells us that the torsion 
$T_{IJK}$ controls how the gauge group is embedded in the global symmetry 
group of the truncated theory. It plays the role of the embedding tensor, 
which is known from truncations that preserve maximal or half-maximal 
supersymmetry (see \cite{Samtleben:2008pe,Trigiante:2016mnt} for reviews).

Eventually, we want to get rid of the internal manifold's coordinate 
dependence. To this end, we note that $K_{\dot\alpha}{}^A = \Kt_{\alphad}{}^I \Et^A{}_I$ in flat indices is constant. The same of course also holds for the torsion $T_{ABC}$ and $T_A$. Thus, we define the constant tensors
\begin{align}
  \eta_{\alphad\betad} &= K_{\alphad}{}^A K_{\betad}{}^B \eta_{AB}\,, \\
  h_{\alphad\betad} &= K_{\alphad}{}^A K_{\betad}{}^B h_{AB}(x)\,, \quad \text{and} \\
  (T_{\alphad})_A{}^B &= \Kt_{\alphad}{}^I T_{IJ}{}^K \Et_A{}^J \Et^B{}_K \,.
\end{align}
Note that $\eta_{\alphad\betad}$ is invariant under the action of the 
gauge group, and we use it to raise and lower dotted Greek indices. 
Moreover, we have to deal with the dilaton. It decomposes into
\begin{equation}
  \phi(x, y) = \phib(x) + \dt(y)\,.
\end{equation}
With these definitions in place, we can eventually restrict the action \eqref{eqn:SNSNSsplit} to $M_{D-n}$,
\begin{equation}
  \begin{aligned}
    S = V_{\mathrm{int}} \int d^{(D-n)} x \, \sqrt{g} e^{-2\phib} \Bigl(& \Rb + 4 \Dc_\mu \phib \Dc^\mu \phib - \frac1{12} \HH_{\mu\nu\rho} \HH^{\mu\nu\rho} \\
    & \frac18 \Dc_\mu h_{AB} \Dc^\mu h^{AB} - \frac14 h_{\alphad\betad} \FF_{\mu\nu}{}^{\alphad} \FF^{\mu\nu\betad} - V \Bigr)\,.
  \end{aligned}
\end{equation}
This reduced action employs the covariant derivative and field strengths
\begin{align}
  \Dc_\mu \phib &= \partial_\mu \phib - \frac12 A_\mu\,, \\ 
  \Dc_\mu h_{AB} &= \partial_\mu h_{AB} - 2 A_\mu{}^{\alphad} (T_{\alphad})_{(A}{}^C h_{B)C}\,,  \\
  \FF_{\mu\nu}{}^{\alphad} &= 2 \partial_{[\mu} A_{\nu]}{}^{\alphad} - f_{\betad\gammad}{}^{\alphad} A_\mu{}^{\betad} A_\nu{}^{\gammad}\,, \qquad \text{and}\\
  \HH_{\mu\nu\rho} &= 3 \partial_{[\mu} B_{\nu\rho]} + 3 A_{[\mu}{}^{\alphad} \partial_\nu A_{\rho]\alphad} - f_{\alphad\betad\gammad} A_\mu{}^{\alphad} A_\nu{}^{\betad} A_\rho{}^{\gammad}\,,
\end{align}
and the gauge transformations
\begin{align}
  \delta_\Lambda A_\mu{}^{\alphad} &= \partial_\mu A_\mu{}^{\alphad} + \Lambda^{\betad} A_\mu{}^{\gammad} f_{\betad\gammad}{}^{\alphad}\,, \\
  \delta_\Lambda \phib &= \frac12 \Lambda^{\alphad} T_{\alphad}\,, \\
  \delta_\Lambda h_{AB} &= 2 \Lambda^{\alphad} (T_{\alphad})_{(A}{}^C h_{B)C}\,.
\end{align}
Two new quantities,
\begin{equation}
  A_\mu (x) = A_\mu{}^{\alphad} (x) T_{\alphad},  \quad \text{and} \quad T_{\alphad} = K_{\alphad}{}^A T_A\,,
\end{equation}
have appeared here. As intended, neither depends on the internal 
coordinates $y$ because the torsion $T_A$ is constant by theorem~\ref{th:CT}. 
There is only one imprint of the manifold $M_d$ left, namely 
its generalised volume
\begin{equation}
  V_{\mathrm{int}} = \int \dd^n y \, e^{-2 \dt}\,,
\end{equation}
which appears as an overall prefactor.

Before, we turn to an example, let us summarise the salient features of 
all truncated theories that arise from theorem~\ref{th:CT} in 
generalised geometry:
\begin{itemize}
  \item They are gauged (super)gravities in dimensions $D < 10$ for the superstring or $D < 26$ for the bosonic string.
  \item Their field content and gauge group are completely fixed by the 
constant, singlet intrinsic torsion $T_{ABC}$ and $T_A$.
  \item The only part of the action that is not so easily fixed is the 
scalar potential $V$. It requires detailed knowledge about the geometry of 
the internal manifold and its dependence on the scalar moduli fields.
\end{itemize}
However, the scalar potential is central for most applications. That is 
the reason why a particular subclass of consistent truncations, called 
generalised Scherk-Schwarz reductions 
\cite{Aldazabal:2011nj,Geissbuhler:2011mx}, currently dominates most 
applications.

\subsection{Generalised Scherk-Schwarz reductions and Poisson-Lie T-duality}\label{sec:gSSandPLTD}
Remarkably, these reductions are directly related to the second central pillar of this paper, dualities. In this subsection, we explain how. 

\subsubsection*{Consistent truncation perspective}
Generalised Scherk-Schwarz reductions implement a special case of 
theorem~\ref{th:CT} with trivial structure group $F$. Therefore, any tensor 
in flat indices which is annihilated by the covariant derivative 
$\nabla_I$ forms part of the truncation. In particular, we do not have 
to deal with $\nabla_\alpha$, introduced in 
\eqref{eqn:definvartensors} and \eqref{eqn:defsinglettorsion}. Moreover, 
the spin connection $\Omega_{IA}{}^B$ in \eqref{eqn:defnabla} vanishes.

This situation is similar to a group manifold in standard geometry, where one can always introduce a flat derivative without curvature. The only difference is that in generalised geometry instead of the Riemann tensor, the generalised Riemann tensor \cite{Jeon:2010rw,Jeon:2011cn,Hohm:2011si}
\begin{equation}
  \RR_{IJKL} = 2 \partial_{[I} \Gamma_{J]KL} + 
2 \Gamma_{[I|ML} \Gamma_{|J]K}{}^M + 
\frac12 \Gamma_{MIJ} \Gamma^M{}_{KL} + (IJ) \leftrightarrow (KL)\,,
\end{equation}
or in flat indices after using \eqref{eqn:GammafromOmega}
\begin{equation}
  \begin{aligned}
    \RR_{ABCD} &= 2 E_{[A}{}^I \partial_I \Omega_{B]CD} + 
   2 \Omega_{[A|C}{}^E \Omega_{B]D}{}_E + 
\frac12 \Omega_{EAB} \Omega^E{}_{CD} \\
    & - F_{AB}{}^E \Omega_{ECD} + (AB) \leftrightarrow (CD)\,,
  \end{aligned}
\end{equation}
has to vanish for trivial $F$. Taking into account the vanishing spin 
connection, we find the generalised affine connection
\begin{equation}
  \Gamma_{IJK} = \partial_I E^A{}_J E_{AK} 
\end{equation}
by using \eqref{eqn:GammafromOmega}. From it, we next compute the 
generalised torsions from \eqref{eqn:gentorsionTIJK},
\begin{equation}
  T_{ABC} = - F_{ABC}\,,
\end{equation}
and from \eqref{eqn:gentorsionTI},
\begin{equation}
  T_A =  F_A\,.
\end{equation}
Hence, the conditions for consistent truncations become
\begin{equation}\label{eqn:Fconst}
  F_{ABC} = \text{const.} \qquad \text{and} \qquad F_A = \text{const.}\,,
\end{equation}
because the covariant derivative $\nabla_A$ acting on quantities with just 
flat indices reduces to $D_A = E_A{}^I \partial_I$. 

At this point, 
it is instructive to look at the Bianchi identities for $\nabla_A$. They reduce to \cite{Geissbuhler:2013uka}
\begin{align}\label{eqn:bianchiFABC}
  D_{[A} F_{BCD]} - \frac34 F_{[AB}{}^E F_{CD]E} &= 0 \,, \quad \text{and} \\\label{eqn:bianchiFA}
  D_{[A} F_{B]} + \frac12 D^C F_{CAB} - \frac12 F^C F_{CAB} &= 0 \,.
\end{align}
Because of \eqref{eqn:Fconst}, all terms with derivatives $D_A$ drop out. 
Consequentially, \eqref{eqn:bianchiFABC} becomes the Jacobi identity of a 
Lie algebra, Lie($\DD$). Assume that this Lie algebra has the generators 
$t_A$, satisfying
\begin{equation}
  [t_A, t_B] = F_{AB}{}^C t_C \,.
\end{equation}
Furthermore, \eqref{eqn:bianchiFA} states that $\tF = F^A t_A$ has to be in the center of Lie($\DD$), saying that the generator $\tF$ commutes with all other elements of the Lie algebra. 

Another effect of a trivial generalised structure group is that we can 
identify the index $\dot\alpha$, enumerating invariant constant vectors, 
with the O(D,D) index $A$, resulting in $K_A{}^B = \delta_A^B$. Hence, 
the Lie group $\DD$ has a natural interpretation as the gauge group of the truncated gauged (super)gravity. In the same vein, we identify the scalar manifold as $O(n,n)/H_n$. Finally, one can directly read off the scalar potential\footnote{We drop the contribution
  \begin{equation}
    F_A F^A - \frac16 F_{ABC} F^{ABC} 
\end{equation}
because it vanishes under the section condition, we always impose in this paper.}
\begin{equation}
  V = F_{ACE} F_{BDF} \left( \frac1{12} h^{AB} h^{CD} h^{EF} - \frac14 h^{AB} \eta^{CD} \eta^{EF} \right) + F_A F_B h^{AB}
\end{equation}
from \eqref{eqn:genRicciScalar} by dropping all terms with a derivative and expanding the projectors $P^{AB} = \frac12 ( \eta^{AB} + h^{AB} )$ and $\Pb^{AB} = \frac12 ( \eta^{AB} - h^{AB} )$.

Summarising, the truncation ansatz for a consistent truncation with a trivial structure group is built from following data:
\begin{enumerate}
  \item A doubled Lie group $\DD$ (=gauge group), generated by $t_A$
    \begin{enumerate}
      \item with the structure coefficients $F_{AB}{}^C$ and
      \item\label{item:FA} and an elements in the center $\tF = F^A t_A$ ($\tF = 0$ always works).
    \end{enumerate}
  \item $\DD$ has to be a subgroup of O($n$,$n$). Otherwise, its 
adjoint action would not leave $\eta_{AB}$ invariant and 
consequently, $F_{ABC} = F_{AB}{}^D \eta_{DC}$ would only be antisymmetric 
with respect to the first two indices $A$ and $B$, but it has to 
be totally antisymmetric. Therefore, $F_{AB}{}^C$ actually describes 
how $\DD$ is embedded into O($n$,$n$), and is called the embedding tensor.
  \item A constant generalised metric $h_{AB}$ on the internal manifold, 
to construct the projectors $P^{AB}$ and $\Pb^{AB}$.
\end{enumerate}

\subsubsection*{Generalised T-duality perspective}
Intriguingly, exactly the same data are needed to describe a 
Poisson-Lie symmetric target space in the $\EE$-model 
formalism \cite{Klimcik:1995dy,Klimcik:2015gba}. More precisely, one 
needs the following ingredients to construct an $\EE$-model:
\begin{enumerate}
  \item A doubled Lie group $\DD$, generated by $t_A$
    \begin{enumerate}
      \item with the structure coefficients $F_{AB}{}^C$.
      \item there is no item~(b).
    \end{enumerate}
  \item A non-degenerate pairing $\langle t_A, t_B \rangle = \eta_{AB}$, that is invariant under the adjoint action of $\DD$.
  \item An $\EE$-operator $\EE: \mathrm{Lie}(\DD) \rightarrow \mathrm{Lie}(\DD)$, which squares to the identity. In the language we use here, this is just the generalised metric $\EE_A{}^B = \HH_A{}^B$.
\end{enumerate}

The underlying classical \s-model 
\begin{equation}\label{eqn:sigmamodel}
  S_\Sigma = \frac{1}{4\pi\alpha'} \int_\Sigma \left(g_{ij} \dd x^i \wedge \star\dd x^j + B_{ij} \dd x^i \wedge \dd x^j \right)
\end{equation}
does not incorporate the dilaton, therefore item \ref{item:FA} is not contained in this list. To define the $\EE$-model, one first transitions to the Hamiltonian formalism with the Hamiltonian\cite{Tseytlin:1990nb}
\begin{equation}
  H = \frac{1}{4\pi\alpha'} \int \dd \sigma \JJ^M \HH_{MN} \JJ^N \,.
\end{equation}
In addition to the generalised metric in \eqref{eqn:genmetric}, we find the 
generalised currents
\begin{equation}
  \JJ_M = \begin{pmatrix} p_m \quad& \partial_\sigma x^m \end{pmatrix}
\end{equation}
defined by using the embedding coordinates $x^m$ and their canonical momenta
\begin{equation}
  p_m = g_{mn} \partial_\tau x^n + B_{mn} \partial_\sigma x^n \,.
\end{equation}
Taking into account canonical, equal-time Poisson brackets $\{x_m(\sigma), p^n(\sigma')\}=\delta_m^n \delta(\sigma-\sigma')$, one obtains
\begin{equation}
  \{ \JJ^M(\sigma), \JJ^N(\sigma') \} = 2\pi\alpha' \delta'(\sigma-\sigma') \eta^{MN}
\end{equation}
which introduces $\eta^{MN}$ from the worldsheet perspective. The $\EE$-models arises after dressing these currents with the generalised frame $E^A{}_I$ to obtain the Kac-Moody algebra
\begin{equation}
  \{ \JJ^A(\sigma), \JJ^B(\sigma') \} = \delta(\sigma-\sigma') F^{AB}{}_C \JJ^C(\sigma) + \delta'(\sigma-\sigma') \eta^{AB}
\end{equation}
with
\begin{equation}
  \JJ^A = \frac{1}{\sqrt{2\pi\alpha'}} E^A{}_M \JJ^M\,.
\end{equation}
As in our previous discussion, it is crucial here that 
$F_{ABC}$, $\eta_{AB}$ and $\HH_{AB}$ are all constant, resulting in 
the $\EE$-model Hamiltonian
\begin{equation}
  H = \frac12 \int \dd \sigma \JJ^A \HH_{AB} \JJ^B
\end{equation}
that is quadratic in the generalised currents and therefore results in 
the simple equations of motion
\begin{equation}
  \dd \JJ + \frac12 [ \JJ, \JJ ] = 0
\end{equation}
with the Lie($\DD$)-valued, worldsheet one-forms
\begin{equation}
  \JJ = t_A \left( \EE^A{}_B \JJ^B \dd\tau + \JJ^A \dd\sigma \right)\,.
\end{equation}

Poisson-Lie T-duality relates different choices for the metric $g_{ij}$ and 
the $B$-field $B_{ij}$ in \eqref{eqn:sigmamodel} by a canonical 
transformation. How this exactly works is best seen in the Hamiltonian 
formalism where the canonical transformation leaves the Poisson brackets 
and the Hamiltonian invariant, but changes the composition of the 
currents $\JJ^M$ in terms of the fundamental field $x^n$. Consequently, there 
is in general not just one choice of the generalised frame $E^A{}_I$ that 
results in some fixed structure constants $F_{AB}{}^C$, but multiple ones.

There is an important lesson to be learned from this new perspective: 
The truncation ansatz, which is fixed by the same generalised frame 
$E_A{}^I$ as the currents $\JJ^A$ in the $\EE$-model, is in general 
not unique. Instead one can find for every maximally isotropic subgroup 
$H_i$ of $\DD$ a generalised frame on the coset $M^{(i)}=H_i \backslash \DD$ 
that results in the same generalised fluxes $F_{ABC}$ 
\cite{Hassler:2017yza,Demulder:2018lmj,Borsato:2021vfy}. All of them are 
connected by Poisson-Lie T-duality. We will discuss the details of 
the construction of generalised frames in section~\ref{sec:vielbeins}.

It is known that gauged $\EE$-models give rise to an even broader notion of 
T-duality \cite{Klimcik:2019kkf}, called dressing cosets 
\cite{Klimcik:1996np}. There are hints that they are also closely related 
to consistent truncations \cite{Demulder:2019vvh}. In the rest of this 
paper, we follow these hints and eventually show that generalised cosets 
provide a very large class of new consistent truncations for which 
the scalar potential can be computed.

\section{The \PS{} construction}
In the last section, we have identified the generalised structure group 
$F$ and the singlet, intrinsic torsion as the fundamental building blocks 
of consistent truncations. We will now present a construction of the 
associated generalised frames $E_A{}^I$ and spin connection 
$\Omega_{AB}{}^C$, which treats them as first class citizens. The basic idea 
for our approach first came up in the paper \cite{Polacek:2013nla}, and we 
therefore refer to it as the \PS{} construction. In its original form, 
it was restricted to the case $F=H_n$, which is not of much use for the 
application to consistent truncations. Fortunately, one of the authors 
extended the discussion to general $F$'s in ref.~\cite{Butter:2022gbc}. 
We shall review the construction in the following, and adapt it to our 
conventions before applying it to truncations.

\subsection{Generalised frame on the mega-space}\label{sec:mega-vielbein}
First, we define generators $t_\alpha\in$ Lie($F$) that generate the generalised structure group and that are governed by the commutators
\begin{equation}\label{eqn:deffalphabetagamma}
  [t_\alpha, t_\beta] = f_{\alpha\beta}{}^\gamma t_\gamma\,.
\end{equation}
Next, we introduce the auxiliary coordinates $z^\mu$ to parameterise 
group elements $f(z^\mu)\in F$. In combination with the coordinates 
$y^i$ on the internal manifold $M_n$, they describe what we call 
mega-space. It is important to keep in mind that the mega-space is 
not physical. It is, rather, a useful book-keeping device, as will become 
clear by the end of this section.

In order to make contact with the discussion in the last section, we have to fix at least two quantities on the mega-space, namely the generalised frame and the $\eta$-metric. For the former, we will use the parameterisation
\begin{equation}\label{eqn:mega-vielbein}
  \Eh_{\Ah}{}^{\Ih} = \Mt_{\Ah}{}^{\Bh} \begin{pmatrix}
    \delta_\beta{}^\gamma & 0 & 0 \\
    -\Omega^\gamma{}_B & E_B{}^J & 0 \\
    \rho^{\beta\gamma} - \frac12 \Omega^\beta{}_K \Omega^{\gamma K} & \Omega^{\beta J} & \delta^\beta{}_\gamma
  \end{pmatrix}
  \begin{pmatrix}
    \vth_\gamma{}^\mu & 0 & 0 \\
    0 & \delta_J^I & 0 \\
    0 & 0 & \vt^\gamma{}_\mu 
  \end{pmatrix}\,.
\end{equation}
Before motivating it, let us take a moment to explain the new conventions we encounter here: Indices come in two kinds: flat $\Ah$, $\Bh$, \dots and curved $\Ih$, $\Jh$, \dots. The latter split as $x^{\Ih} = \begin{pmatrix} z^\mu & y^i & \yt_i & \zt_\mu \end{pmatrix}$, $x_{\Ih} = \begin{pmatrix} \zt_\mu & \yt_i & y^i & z^\mu \end{pmatrix}$. Note that the section condition is still trivially solved, because nothing depends on the coordinates $\yt_i$ and $\zt_\mu$ and we use the partial derivative $\partial_{\Ih} = \begin{pmatrix} \partial_\mu & \partial_i & 0 & 0 \end{pmatrix}$. Moreover, the O($n+\dim F$,$n+\dim F$) invariant metric $\eta_{\Ih\Jh}$ is given by
\begin{equation}
  \eta_{\Ih\Jh} = \begin{pmatrix}
    0 & 0 & \delta_\mu^\nu \\
    0 & \eta_{IJ} & 0 \\
    \delta_\nu^\mu & 0 & 0
  \end{pmatrix}
\end{equation}
where $\eta_{IJ}$ is already known from \eqref{eqn:etametric}. 
Flat indices behave in the same way. In particular, we encounter the flat $\eta$-metric
\begin{equation}
  \eta_{\Ah\Bh} = \begin{pmatrix}
    0 & 0 & \delta_\alpha^\beta \\
    0 & \eta_{AB} & 0 \\
    \delta_\beta^\alpha & 0 & 0
  \end{pmatrix}\,.
\end{equation}

 With the index conventions established, we can say more about the 
form of \eqref{eqn:mega-vielbein}. At first glance, it has three 
distinguishing features:
\begin{enumerate}
  \item All physically relevant quantities are contained in the middle matrix.
  \item This matrix is of lower triangular form.
  \item It is dressed from left and right by two matrices which only 
depend on the auxiliary coordinates $z^\mu$.
\end{enumerate}
Note that we use here tilde quantities which depend on the auxiliary coordinates 
$z^\mu$. $\Mt_{\Ah}{}^{\Bh}$ does not appear explicitly
in either \cite{Polacek:2013nla} or \cite{Butter:2022gbc}, 
but later it will be very helpful for understanding how the 
mega-generalised frame 
field 
relates to the covariant derivative $\nabla_A$. This matrix mediates 
the adjoint action of $F$ on the mega-space. Therefore, it has two 
defining properties:
\begin{equation}\label{eqn:MtODD}
  \Mt_{\Ah}{}^{\Ch} \Mt_{\Bh}{}^{\Dh} \eta_{\Ch\Dh} = \eta_{\Ah\Bh}
\end{equation}
and
\begin{equation}\label{eqn:Mtaction}
  \partial_\mu \Mt_{\Ah}{}^{\Bh} = \vt^\alpha{}_\mu \Mt_{\Ah}{}^{\Ch} f_{\alpha \Ch}{}^{\Bh}\,.
\end{equation}
Here $\vt^\alpha{}_\mu$ denotes the components of the right-invariant 
one-forms $\dd f f^{-1} = t_\alpha \vt^\alpha{}_\mu \dd z^\mu$ and 
$\vth_\alpha{}^\mu$ are the dual vector fields, defined by 
$\vth_\alpha{}^\mu \vh^\beta{}_\mu = \delta_\alpha^\beta$. The 
infinitesimal action of $F$ is specified by the constants 
$f_{\alpha\Bh\Ch}$. Due to \eqref{eqn:MtODD},  they have to satisfy 
$f_{\alpha\Bh\Ch} = -  f_{\alpha\Ch\Bh}$. Furthermore, 
\eqref{eqn:Mtaction} should not spoil the lower triangular form of the 
middle matrix in \eqref{eqn:mega-vielbein} and therefore we find
\begin{equation}
  f_{\alpha\beta}{}^C = 0 \,, \qquad \text{and} \qquad
  f_{\alpha\beta\gamma} = 0 \,.
\end{equation}
Owing to these constraints, the full mega-generalised frame field is 
lower triangular. This observation motivates the parameterisation of the 
two right-most matrices in \eqref{eqn:mega-vielbein}. Together, they 
implement the most general lower triangular matrix that leaves the 
$\eta$-metric invariant. Thereby, we encounter three fields on the 
physical space $M_n$:
\begin{itemize}
  \item The generalised frame $E_A{}^I$ that 
we already discussed in the last section.
  \item An unconstrained field $\Omega^\alpha{}_B$ which, as we will see later, is directly related to the spin connection $\Omega_{ABC}$ of $\nabla_A$.
  \item An antisymmetric tensor $\rho^{\alpha\beta} = - \rho^{\beta\alpha}$. 
It is the most interesting result of the \PS{} construction because it has 
no analog in standard geometry. Therefore, in honor of these two gentlemen,
we call it the \PS{} field.
\end{itemize}

Finally, we come to the question: How is this setup related to the 
covariant derivative $\nabla_A$? The relation becomes manifest when we 
split all tensors into a physical and an auxiliary part as, for example, in
\begin{equation}\label{eqn:Mtsplitting}
  \Vh_{\Ah} = \Mt_{\Ah}{}^{\Bh} V_{\Bh} \,.
\end{equation}
To indicate that $\Vh_{\Ah}$ depends on $z^\mu$ and $y^i$ 
while $V_{\Ah}$ depends only on $y^i$, we decorate the former with a 
hat. This is also the reason that the generalised frame field on the 
mega-space is called $\Eh_{\Ah}{}^{\Ih}$ instead of just $E_{\Ah}{}^{\Ih}$. 
In the same vein, we introduce the flat derivative
\begin{equation}
  \Dh_{\Ah} = \Eh_{\Ah}{}^{\Ih} \partial_{\Ih} = \Mt_{\Ah}{}^{\Bh} D_{\Ah}\,,
    \qquad \text{with} \qquad
  D_{\Ah} = E_{\Ah}{}^{\Ih} \partial_{\Ih}\,.
\end{equation}
A natural question at this point is:  Can get rid of the auxiliary 
coordinates $z^\mu$ completely? This would be the case if these coordinates 
always appeared only in the form of \eqref{eqn:Mtsplitting}. There is only 
one place where this could go wrong, namely for flat derivatives. 
But fortunately, for them we find
\begin{equation}\label{eqn:Dhtonabla'1}
  \Dh_{\Ah} \Vh_{\Bh} = \Mt_{\Ah}{}^{\Ch} \Mt_{\Bh}{}^{\Dh} \left( D_{\Ch} V_{\Dh} + E_{\Ch}{}^\alpha f_{\alpha \Dh}{}^{\Eh} V_{\Eh} \right) \quad \text{with} \quad
  E_{\Ah}{}^{\beta} = \begin{pmatrix}
    \delta_\alpha^\beta \\
    - \Omega^\beta{}_A \\
    \rho^{\alpha\beta} - \frac12 \Omega^{\alpha}{}_I \Omega^{\beta I}
  \end{pmatrix}\,.
\end{equation}
Hence, we conclude that the $z^\mu$-dependence of any quantity with 
flat indices arises only from the twist of each index with $\Mt_{\Ah}{}^{\Bh}$.

Looking more closely at \eqref{eqn:Dhtonabla'1}, one can interpret the 
two terms 
in the brackets on the right hand side of the first equation as a 
covariant derivative. Thus, we write
\begin{equation}\label{eqn:Dhtonabla'2}
  \Dh_{\Ah} \Vh_{\Bh} = \Mt_{\Ah}{}^{\Ch} \Mt_{\Bh}{}^{\Dh} \nabla'_{\Ch} V_{\Dh}
\end{equation}
for flat derivatives on the mega-space, and we see that it can 
alternatively be interpreted as a covariant derivative on $M_n$. 
Comparing \eqref{eqn:Dhtonabla'1} and \eqref{eqn:Dhtonabla'2}, we find
\begin{align}\label{eqn:nabla'A}
  \nabla'_A &= E_A{}^I \partial_I - \Omega^\beta{}_A \nabla'_\beta \,, \\
  \nabla'^\alpha &= \Omega^{\alpha B} D_B + \left( \rho^{\alpha\beta} - \frac12 \Omega^{\alpha}{}_I \Omega^{\beta I} \right) \nabla'_\beta\,.
\end{align}
and furthermore
\begin{align}
  \nabla'_\alpha V_\beta &= f_{\alpha\beta}{}^\gamma V_\gamma \,, \\
  \nabla'_\alpha V_B  &= f_{\alpha B}{}^C V_C + f_{\alpha B}{}^\gamma V_\gamma \,,\\
  \nabla'_\alpha V^\beta &= - f_{\alpha\gamma}{}^\beta V^\gamma - f_{\alpha C}{}^\beta V^C + f_\alpha{}^{\beta\gamma} V_\gamma\,.
\end{align}
Most important for our purpose is to relate $\nabla'_{\Ah}$ to the 
covariant derivative $\nabla_A$ and $\nabla_\alpha$ that is required 
for constructing consistent truncations. More precisely, we want to identify
\begin{equation}\label{eqn:nablaAtonabla'A}
  \nabla_A V_B = \nabla'_A V_B\,,
\end{equation}
or equally
\begin{equation}
  \Omega_{IBC} = \Omega^\delta{}_I f_{\delta BC}\,.
\end{equation}
Hence, we impose
\begin{equation}
  f_{\alpha B}{}^\gamma = 0\,.
\end{equation}
This condition can be relaxed in the context of generalised 
T-dualities\footnote{We thank Yuho Sakatani for pointing out this 
possibility to us.}. However, for the consistent truncations we study here, 
it arises naturally. Thus, we shall keep the 
$\cancel{f_{\alpha B}{}^\gamma}$-terms in 
intermediate results and only remove them in the final expression on $M_n$.

Finally, we also need to know how the generalised structure group $F$ acts on the physical generalised tangent space $T M_n \oplus T^* M_n$. Remember, the corresponding infinitesimal action is mediated by $\nabla_\alpha$, introduced in \eqref{eqn:definvartensors}. Thus, it is natural to relate
\begin{equation}\label{eqn:nablaalphatonabla'alpha}
  \nabla_\alpha V_B := f_{\alpha B}{}^C V_C = \nabla'_\alpha V_B \,,
\end{equation}
too. This also explains the initially arbitrary looking notation for this operation. On the mega-space it has exactly the same origin as the covariant derivative $\nabla_A$.

\subsection{Torsion and curvature}\label{sec:torsionandcurvature}

We have seen that the \PS{} construction transforms flat derivatives 
$\Dh_{\Ah}$ on the mega-space into covariant derivatives on the physical 
space $M_n$. From a conceptual point of view, one might say that it geometrises 
a generalised connection by adding auxiliary coordinates $z^\mu$. Therefore, 
analysing the properties of the derivatives $\Dh_{\Ah}$ becomes the 
main objective for this subsection. Fortunately, we already learned in 
section \eqref{sec:gSSandPLTD} that they are exclusively controlled by 
the torsions (see \eqref{eqn:FABC} and \eqref{eqn:FA})
\begin{align}\label{eqn:fhABC}
  \fh_{\Ah\Bh\Ch} &= \Dh_{[\Ah} \Eh_{\Bh}{}^{\Ih} \Eh_{\Ch]\Ih} \,, \quad \text{and} \\
  \fh_{\Ah} &= 2 \Dh_{\Ah} \dh - \partial_{\Ih} \Eh_{\Ah}{}^{\Ih} \,.
\end{align}
Here, we switched from capital $F$'s to $f$'s, because we want to reserve 
the $F$ for the physical space $M_n$. At the moment this might seem 
arbitrary, but it will become obvious shortly. Next, we will evaluate 
\eqref{eqn:fhABC}. As already discussed, it is convenient to strip off 
$\Mt_{\Ah}{}^{\Bh}$ from generalised tensors, as we did in 
\eqref{eqn:Mtsplitting} while going from $\Vh_{\Ah}$ to $V_{\Ah}$. 
The generalised frame field in \eqref{eqn:mega-vielbein} is no exception. 
Therefore, we split it according to
\begin{equation}
  \Eh_{\Ah}{}^{\Ih} = \Mt_{\Ah}{}^{\Bh} \Eb_{\Bh}{}^{\Ch} \VV_{\Ch}{}^{\Ih}
    \qquad \text{with} \qquad
  \VV_{\Ah}{}^{\Ih} = \begin{pmatrix}
    \vth_\alpha{}^\mu & 0 & 0 \\
    0 & E_A{}^I & 0 \\
    0 & 0 & \vt^\alpha{}_\mu
  \end{pmatrix}\,,
\end{equation}
and first compute
\begin{equation}
  3 \VV_{[\Ah}{}^{\Ih} \partial_{\Ih} \VV_{\Bh}{}^{\Jh} \VV_{\Ch]\Jh} = \begin{cases}
    f_{\alpha\beta}{}^{\gamma} \text{ and cyclic} \\
    F_{ABC}\,.
  \end{cases}
\end{equation}
The latter is then used to obtain
\begin{equation}
  f_{\Ah\Bh\Ch} = 3 \Eb_{\Ah}{}^{\Dh} \Eb_{\Bh}{}^{\Eh} \Eb_{\Ch}{}^{\Fh} \VV_{[\Dh}{}^{\Ih} 
  \partial_{\Ih} \VV_{\Eh}{}^{\Jh} \VV_{\Fh]\Jh} + 3 \nabla'_{[\Ah} \Eb_{\Bh}{}^{\Ih} \Eb_{\Ch]\Ih}\,.
\end{equation}
Note that $\nabla'_A$ here just acts on the flat index $\Bh$ of the 
generalised frame field $\Eb_{\Bh}{}^{\Ih}$. There is no affine 
connection fixed by a vielbein postulate. Before we can turn to $\fh_{\Ah}$, 
we have to specify how the generalised dilaton on the mega-space depends 
on the auxiliary coordinates $z^\mu$. It turns out that the right choice is
\begin{equation}
  \dh(y, z) = d(y) - \frac12 \log \det \et(z)
\end{equation}
with
\begin{equation}
  t_\alpha \et^\alpha{}_\mu(z) \dd z^\mu = f^{-1} \dd f\,.
\end{equation}
Only for this choice does 
the $z^\mu$-dependence of $\fh_{\Ah}$ completely factor 
into a $\Mt_{\Ah}{}^{\Bh}$ twist, as we required in \eqref{eqn:Mtsplitting}. 
After removing this twist, we find
\begin{equation}\label{eqn:fhA}
  f_{\Ah} = 2 \Eb_{\Ah}{}^B D_B d - \partial_I E_{\Ah}{}^I - f_{\alpha\Ah}{}^{\Bh} \Eb_{\Bh}{}^\alpha\,.
\end{equation}

It is very instructive to compute the individual components of 
both $\fh_{\Ah\Bh\Ch}$ and $\fh_{\Ah}$. But before doing so, we need a 
way to keep track of all contributions. To this end, we introduce the 
$\epsilon$-dimension, which is defined in the following way: Assume we 
scale the generators of the generalised structure group $F$ according to 
$t_\alpha \rightarrow \epsilon^{-1} \, t_\alpha$. In this case, the 
structure coefficients $f_{\alpha\beta}{}^\gamma$ introduced in 
\eqref{eqn:deffalphabetagamma} scale as 
$f_{\alpha\beta}{}^\gamma \rightarrow \epsilon^{-1} 
f_{\alpha\beta}{}^\gamma$. To find out how other tensors scale, 
assign $-1$ to each lowered Greek index, $+1$ to each raised Greek index 
and $0$ to each Latin index. Summing over all indices of the tensor
then gives its $\epsilon$-dimension. The motivation for this particular 
scaling comes from an alternative approach in the literature to 
the construction of dressing cosets \cite{Sfetsos:1999zm,Sakatani:2021skx}. 
It considers an $\EE$-model on the mega-space where the generalised 
structure group $F$ describes a global symmetry. After scaling the 
generators $t_\alpha$ as shown above, and sending $\epsilon$ to $0$, 
the $\EE$-model degenerates and the global symmetry becomes a gauge 
symmetry. This limit is subtle, but several examples suggest that the 
relevant quantities on the dressing coset are those which are invariant 
under the scaling or, equally, have vanishing $\epsilon$-dimension.

To find all independent components of $\fh_{\Ah\Bh\Ch}$, recall
that it is by construction totally antisymmetric. For each of its 
components, we can therefore order the indices by their 
$\epsilon$-dimension (of course still keeping track of the sign). 
The results for the ten independent classes of components are then given
by
\begin{equation}\label{eqn:decompfhABC}
  \begin{tabular}{r|ll|l}
    $\epsilon$-dim. & & & \\
    $-3$ & $f_{\alpha\beta\gamma} = 0$ & & \\
    $-2$ & $f_{\alpha\beta C} = 0$ &  & \\
    $-1$ & $f_{\alpha\beta}{}^\gamma$ & $f_{\alpha AB}$ & \\
     $0$ & $f_{\alpha B}{}^\gamma = 0$ & & $f_{ABC}$ \\
    $+1$ & $f_{\alpha}{}^{\beta\gamma}$ & & $f_{AB}{}^\gamma$ \\
    $+2$ & & & $f_A{}^{\beta\gamma}$ \\
    $+3$ & & & $f^{\alpha\beta\gamma}$\,.
  \end{tabular}
\end{equation}
All the components in the first column are just the parts of 
$f_{\alpha \Bh}{}^{\Ch}$ that describe the $F$-action on the mega-space 
(see \eqref{eqn:Mtaction}). They are constant, and constrained by the 
Jacobi identity
\begin{equation}\label{eqn:closureFaction}
  2 f_{[\alpha|\Ch}{}^{\Eh} f_{|\beta]\Eh}{}^{\Dh} = - f_{\alpha\beta}{}^\gamma f_{\gamma\Ch}{}^{\Dh}\,,
\end{equation}
which arises from $\dd^2 M_{\Ah}{}^{\Bh}=0$ and \eqref{eqn:Mtaction}, 
because infinitesimal $F$-transformations have to close into a group. 
Non-constant contributions only arise from the components in the second column. In the following, we compute them one by one.

First, we look at
\begin{equation}
  f_{ABC} = F_{ABC} - 3 \Omega^\delta{}_{[A} f_{\delta BC]} = F_{ABC} - 3 \Omega_{[ABC]} \,.
\end{equation}
Comparing this result with \eqref{eqn:GammafromOmega} and \eqref{eqn:gentorsionTIJK}, we find the remarkable identification
\begin{equation}
  f_{ABC} = - T_{ABC}\,.
\end{equation}
On the other hand, \eqref{eqn:defsinglettorsion} requires that $T_{BCD}$ is annihilated by $\nabla_A$ and $\nabla_\alpha$ for theorem~\ref{th:CT} to apply. This condition can now be written as
\begin{align}\label{eqn:nablaAfBCD}
  \nabla_A f_{BCD} = \nabla'_A f_{BCD} &= 0 \quad \text{and} \\
  \nabla_\alpha f_{BCD} = \nabla'_{\alpha} f_{BCD} &=0
\end{align}
by using \eqref{eqn:nablaAtonabla'A} and \eqref{eqn:nablaalphatonabla'alpha}. Due to \eqref{eqn:nabla'A}, these two equations automatically imply
\begin{equation}\label{eqn:fABCconst}
  f_{ABC} = \text{const.}
\end{equation}
for all consistent truncations resulting from theorem~\ref{th:CT}. Next in line is
\begin{equation}\label{eqn:fABgamma}
  f_{AB}{}^\gamma = - 2 D_{[A} \Omega^\gamma{}_{B]} + f_{\alpha\beta}{}^\gamma \Omega^\alpha{}_A \Omega^\beta{}_B - \frac12 f_{\alpha A B} \Omega^\alpha{}_C \Omega^{\gamma C} + \cancel{2 f_{\alpha[A}{}^\gamma \Omega^\alpha{}_{B]}} - f_{\alpha AB}\rho^{\alpha\gamma} + F_{AB}{}^C \Omega^\gamma{}_C \,.
\end{equation}
To find an interpretation for this quantity, we first get rid of the Greek index $\gamma$ by contracting with $f_{\gamma CD}$, resulting in the new quantity
\begin{equation}\label{eqn:deffABCD}
  f_{ABCD} := f_{AB}{}^\gamma f_{\gamma CD}\,.
\end{equation}
In the same vein, we introduce
\begin{equation}
  r_{ABCD} := \rho^{\alpha\beta} f_{\alpha AB} f_{\beta CD}\,,
\end{equation}
which is actually the original form of the \PS-field 
introduced in \cite{Polacek:2013nla}. With these two new quantities, 
\eqref{eqn:fABgamma} can be rewritten exclusively in terms of 
Latin indices as
\begin{equation}
  f_{ABCD} = - 2 D_{[A} \Omega_{B]CD} - 2 \Omega_{[A|C}{}^E \Omega_{B]DE} - \frac12 \Omega_{EAB} \Omega^E{}_{CD} + F_{AB}{}^E \Omega_{ECD} - r_{ABCD}\,.
\end{equation}
From \eqref{eqn:deffABCD}, it follows that $f_{ABCD}$ is antisymmetric with respect to its first two and last two indices. Thus, it can be decomposed into the three contributions
\begin{equation}\label{eqn:decompfABCD}
  f_{ABCD} = \ydiagram{1,1} \otimes \ydiagram{1,1} = \ydiagram{1,1,1,1} \oplus \ydiagram{2,2} \oplus \ydiagram{2,1,1} \,.
\end{equation}
The first two diagrams on the right hand side contain the parts 
of $f_{ABCD}$ that are symmetric under the pairwise exchange of 
$AB$ and $CD$. For them, $r_{ABCD}$ drops out and we find
\begin{equation}\label{eqn:ftoRR}
  f_{ABCD} + f_{CDAB} = - \RR_{ABCD}\,.
\end{equation}
At this point, the power of the \PS{} construction becomes fully apparent: 
We recover the generalised curvature on the physical space $M_n$ purely
from a torsion component on the mega-space!

Additionally, the Bianchi identity components  
\begin{equation}\label{eqn:BIfhAhBhCh}
  \Dh_{[A} \fh_{BCD]} - \frac34 \fh_{[AB}{}^{\Eh} \fh_{CD]\Eh} = 0
\end{equation}
on the mega-space imply the algebraic Bianchi identity of $\RR_{ABCD}$, 
namely
\begin{equation}\label{eqn:algBIgenRiem}
  \RR_{[ABCD]} = f_{[AB}{}^E f_{CD]E} - \frac43 \nabla_{[A} f_{BCD]}\,.
\end{equation}
Note that $\nabla_A f_{BCD} = 0$ for consistent truncations, rendering the 
identity completely algebraic. Finally, there is the hook in 
\eqref{eqn:decompfABCD}, which is antisymmetric under pairwise 
exchange of $AB$ and $CD$ and contains the \PS-field. As we will show in 
the next subsection, it will not contribute to the two-derivative action 
or its equations of motion. The same holds for the remaining components 
$f_A{}^{\beta\gamma}$ and $f^{\alpha\beta\gamma}$. Hence, we postpone 
deriving detailed expressions for them until section~\ref{sec:higherderiv}.

  We still have to deal with the one-index torsion \eqref{eqn:fhA}. 
In analogy with \eqref{eqn:decompfhABC}, we decompose it into
\begin{equation}\label{eqn:complistfAh}
  \begin{tabular}{r|l|l}
    $\epsilon$-dim. & & \\
    $-1$ & $f_\alpha = f_{\alpha\beta}{}^\beta$ & \\
     $0$ & & $f_A$ \\
     $1$ & & $f^\alpha$\,.
  \end{tabular}
\end{equation}
Again, we need to compute the two contributions, $f_A$ and $f^\alpha$, 
in the right-hand column.  For the first, we obtain
\begin{equation}
  f_A = F_A - \Omega^B{}_{BA} - \cancel{f_{\beta A}{}^\beta}
\end{equation}
by expanding \eqref{eqn:fhA}. Similar to $f_{ABC}$, we recover the
one-index torsion for the last section and identify
\begin{equation}
  T_A = f_A
\end{equation}
Therefore,
\begin{equation}
  \nabla_A T_B = \nabla'_A f_B = 0  \qquad \text{and} \qquad \nabla_\alpha T_B = \nabla'_\alpha f_B = 0
\end{equation}
hold according to \eqref{eqn:defsinglettorsion} for all consistent 
truncations that are governed by theorem~\ref{th:CT}. Just as  
in \eqref{eqn:fABCconst}, we therefore deduce 
\begin{equation}\label{eqn:fAconst}
  f_A = \text{const.}
\end{equation}
Finally, there is
\begin{equation}
  f^\alpha = - D_B \Omega^\alpha{}_B - \cancel{\Omega^\beta{}_C f_\beta{}^{C \alpha}} - \rho^{\beta\gamma} f_{\beta\gamma}{}^\alpha + \Omega^{\alpha B} F_B + f_\beta{}^{\beta\alpha}\,,
\end{equation}
which we rewrite as
\begin{align}
  f_{AB} :&= \left( f^\alpha - f_\gamma{}^{\gamma\alpha} \right) f_{\alpha AB} \\
          &=- D^C \Omega_{CAB} + F^C \Omega_{CAB} + 2 r^C{}_{ABC} \,.
\end{align}
However, it is not a new independent quantity. Instead it can be
reconstructed from already  encountered $f$'s by using the Bianchi
identity 
\begin{equation}
  \Dh_{[A} \fh_{B]} + \frac12 \Dh^{\Ch} \fh_{\Ch AB} - \frac12 \fh^{\Ch} \fh_{\Ch AB} = 0
\end{equation}
on the mega-space, resulting in
\begin{equation}\label{eqn:fABfromrest}
  f_{AB} = 2 \nabla_{[A} f_{B]} + \nabla^C f_{CAB} - f^C f_{CAB} - 2 f^C{}_{[AB]C} - f_\alpha f^{\alpha}{}_{AB} \,.
\end{equation}

\subsection{Truncation of the generalised Ricci scalar/tensor and 
dualities}\label{sec:truncgenRicci}

We now come to the core objective of this paper, and reveal the relation 
between consistent truncations and dualities. To do so, we revisit the 
generalised Ricci scalar in \eqref{eqn:genRicciScalar}. Currently, it is 
written in terms of the flat derivative $D_A$. But we rather want to 
write it in terms of the covariant derivatives $\nabla_A$ that are relevant 
for consistent truncations. While $F_A$ and $F_{ABC}$ are directly related 
to the torsion of $D_A$, they lack this kind of geometric interpretation 
for $\nabla_A$. Therefore, we will ultimately replace them with 
$f_A$, $f_{ABC}$, and $f_{ABCD}$. In general, one might worry that 
after performing these substitutions, some naked connection terms would
remain. By naked, we mean any generalised spin connection 
$\Omega_{AB}{}^C$ which is not part of a corresponding covariant 
derivative $\nabla_A$. However, the generalised Ricci scalar transforms 
covariantly under the action of the generalised structure group $F$. 
Therefore, one should in the end be able to rewrite it just in terms of 
manifestly covariant quantities, and these are exactly all the $f$'s 
introduced above (for more details 
see section~\ref{sec:gaugetransformations}) and their covariant derivatives. 
Indeed, all terms including a naked connection vanish and one finds
\begin{equation}
  \RR = P^{AB} P^{CD} \left[ \left( \Pb^{EF} + \frac13 P^{EF} \right) f_{ACE} f_{BDF} + 2 f_{ACDB} \right] + 2 P^{AB} ( 2 \nabla_A f_B - f_A f_B )\,.
\end{equation}
Note that the \PS-field gets projected out from $f_{ABCD}$ and we can also write
\begin{equation}\label{eqn:defRnabla}
  2 P^{AB} P^{CD} f_{ACBD} = - \RR_{ACBD} P^{AB} P^{CD} := \RR^\nabla
\end{equation}
by using \eqref{eqn:ftoRR}. Hence, can rewrite $\RR$ purely in terms of 
the torsions and scalar curvature of $\nabla_A$ as
\begin{equation}\label{eqn:RRfornablaA}
  \RR = \RR^\nabla + P^{AB} P^{CD} \left( \Pb^{EF} + 
\frac13 P^{EF} \right) T_{ACE} T_{BDF} + 
2 P^{AB} ( 2 \nabla_A T_B - T_A T_B )\,.
\end{equation}

As shown in diagram~\eqref{diag:contrunc}, it is crucial for any 
consistent truncation to preserve the relation between the action and 
its equations of motion. Therefore, we compute the latter by varying the 
action \eqref{eqn:SDFT} and dropping all boundary terms, to obtain
\begin{equation}\label{eqn:eomG}
  \delta S = \int \dd^D x\, e^{-2 d} \left( - 2 \RR \delta d + \GG_{AB} \delta E^{AB} \right) = 0\,,
\end{equation}
with \cite{Geissbuhler:2013uka}
\begin{equation}\label{eqn:GAB}
  \begin{aligned}
    \GG_{AB} = 4 P_{[A}{}^C \Pb_{B]}{}^D \Bigl( &F_{CEG} F_{DFH} P^{EF} \Pb^{GH} + F_{CDE} F_F P^{EF} + \\
      & D_D F_C - D_E F_{CDF} P^{EF} \Bigr)
  \end{aligned}
\end{equation}
and
\begin{equation}
  \delta E_{AB} := (\delta E_A{}^I ) E_{BI} \,.
\end{equation}
Note that $\GG_{AB}$ is not yet the generalised Ricci tensor 
$\RR_{IJ}$ that arises if we vary with respect to the generalised 
metric $\mathcal{H}_{IJ}$
\begin{equation}\label{eqn:eomR}
  \delta S = \int \dd^D x\, e^{-2 d} \left( - 2 \RR \delta d + \RR_{IJ} \delta \HH^{IJ} \right) = 0\,.
\end{equation}
Both $\GG_{AB}$ and $\RR_{IJ}$ are commonly used in the literature. They 
can be related by identifying
\begin{equation}
  \delta \HH^{IJ} = 4 \delta E_{KL} P^{K(I} \Pb^{J)L} \qquad \text{and} \qquad
  \delta E^{AB} = \delta \HH_{CD} P^{C[A} \Pb^{B]D}
\end{equation}
first, and afterwards comparing \eqref{eqn:eomG} with \eqref{eqn:eomR} to eventually obtain
\begin{equation}
  \RR_{IJ} = \GG^{LK} P_{L(I} \Pb_{J)K} \,.
\end{equation}

We follow the same strategy as we previously did for for $\RR$,
and rewrite $\GG_{AB}$ in terms of $\nabla_A$, $f_A$, $f_{ABC}$ 
and $f_{ABCD}$. Again, we find that all terms with naked connections 
cancel. The result
\begin{equation}
  \begin{gathered}
    \GG_{AB} = 4 P_{[A}{}^C \Pb_{B]}{}^D \Bigl( f_{CEG} f_{DFH} P^{EF} \Pb^{GH} + f_{CDE} f_F P^{EF} \\
    + \nabla_D f_C - \nabla_E f_{CDF} P^{EF} - f_{EDCF} P^{EF} \Bigr)
  \end{gathered}
\end{equation}
again looks similar to \eqref{eqn:GAB}, with only one new contribution 
from $f_{ABCD}$. If we look at the definition \eqref{eqn:deffABCD}, we note that the last two indices of $f_{ABCD}$ originate from $f_{\alpha CD}$ which captures the infinitesimal action of the generalised structure group $F$ on the generalised tangent space. Because $F$ is a subgroup of the double Lorentz group $H_D$, we find \begin{equation}
  f_{\alpha CD} P^C{}_A \Pb^D{}_B = f_{\alpha CD} \Pb^C{}_A P^D{}_B = 0 \,.
\end{equation}
Exploiting this property, we rewrite
\begin{equation}\label{eqn:defGnablaAB}
  - 4 P_{[A}{}^C \Pb_{B]}{}^D f_{EDCF} P^{EF} = 2 P_{[A}{}^C \Pb_{B]}{}^D \RR_{ECDF} P^{EF} = \GG^\nabla_{AB}
\end{equation}
and see that only the box diagram in the decomposition 
\eqref{eqn:decompfABCD} of $f_{ABCD}$ contributes to the generalised Ricci 
tensor. Like for $\RR$ in \eqref{eqn:RRfornablaA}, we can write it 
exclusively in terms of $\nabla_A$'s torsions and curvature:
\begin{equation}\label{eqn:RRABfornablaA}
  \begin{aligned}
    \GG_{AB} = \GG^\nabla_{AB} + 4 P_{[A}{}^C \Pb_{B]}{}^D \Bigl( & T_{CEG} T_{DFH} P^{EF} \Pb^{GH} - T_{CDE} T_F P^{EF} + \\
    & \nabla_D T_C + \nabla_E T_{CDF} P^{EF} \Bigr)\,.
  \end{aligned}
\end{equation}

For any consistent truncation governed by theorem~\ref{th:CT}, $\RR$, $\RR_{AB}$ and $\GG_{AB}$ have to be constant and singlets. Written in terms of equations this imposes
\begin{equation}
  \nabla_{A/\alpha} \RR = 0 \qquad \text{and} \qquad
  \nabla_{A/\alpha} \GG_{BC} = 0 \,.
\end{equation}
Because the intrinsic torsions $f_A$ and $f_{ABC}$ share this property, the two new quantities $\RR^\nabla$ and $\GG^\nabla_{AB}$ defined in \eqref{eqn:defGnablaAB} and \eqref{eqn:defRnabla} have to satisfy 
\begin{equation}\label{eqn:constraints2deriv}
  \nabla_{A/\alpha} \RR^\nabla = 0 \qquad \text{and} \qquad 
  \nabla_{A/\alpha} \GG^\nabla_{AB} = 0\,,
\end{equation}
too. This fixes some of the components of $F_{\Ah\Bh\Ch}$, which we need for 
the \PS{} construction, to be constants, but clearly not all of them.

However, assuming that we have fixed all physically relevant data 
(at least at the leading two derivative level), we should rather ask: 
Is there a way to choose the remaining components of $f_{\Ah\Bh\Ch}$ 
such that they are compatible with the ones we already fixed? From a 
physical point of view this corresponds to obtaining a truncation ansatz 
which reproduces a target truncated theory. In general, there can be 
multiple solutions to this problem. Each of them provides a different 
metric and $B$-field on the internal space, but eventually they all 
reproduce the same truncated theory. We encountered this behaviour, which 
is depicted in \eqref{eqn:dualitiesdiagram}, already, for the 
motivating example of generalised Scherk-Schwarz reductions and their 
relation to Poisson-Lie T-duality. But a main difference is that we now 
have two possible mechanisms for generating dualities instead of one. 
First, we will see that there is the possibility of finding different 
generalised frames on the mega-space that realise the same generalised 
fluxes $f_{\Ah\Bh\Ch}$.  We already encountered this 
mechanism in section~\ref{sec:gSSandPLTD}; it provides the foundation of 
Poisson-Lie T-duality. Moreover, for a non-trivial generalised structure 
group, we have a second option since only some components of 
$f_{\Ah\Bh\Ch}$ are fixed. The remaining ones are only constrained by 
the Bianchi identities on the mega-space. 

\subsection{Jacobi identities}\label{sec:Jac}

Exploring the space of all these dual backgrounds is beyond the 
scope of the present paper. We rather want to restrict our discussion 
to a specific family of backgrounds with constant $f_{\Ah\Bh\Ch}$, 
like for generalised Scherk-Schwarz reductions. 
In section~\ref{sec:vielbeins}, we show that all of them are realised in 
terms of generalised cosets. Thus, they 1) form the foundation of all 
generalised T-dualities currently known, and 2) provide explicit constructions
of the full truncation ans\"atze. Hence, the question from above has to be refined: Can we find for any given truncated theory at least one ansatz with constant $f_{\Ah\Bh\Ch}$ that uplifts the theory to $D$ dimensions? 

To answer this question, one has to analyse the Bianchi identities
\begin{align}\label{eqn:bianchiFhABC}
  \fh_{[\Ah\Bh}{}^{\Eh} \fh_{\Ch\Dh]\Eh} &= 0 \quad \text{and} \\\label{eqn:bianchiFhA}
  \fh^{\Ch} \fh_{\Ch\Ah\Bh} &= 0\,,
\end{align}
on the mega-space, which arise after dropping the flat derivative 
$\Dh_{\Ah}$ because we are assuming that $f_{\Ah\Bh\Ch}$ and $f_{\Ah}$ are 
constant. It turns out that this problem is still hard, because it 
entails solving systems of coupled quadratic equations. Therefore, we 
postpone a thorough analysis to future work. But even without a full 
solution, already the structure of the various components of 
\eqref{eqn:bianchiFhABC} and \eqref{eqn:bianchiFhA} is interesting.

We recognise \eqref{eqn:bianchiFhABC} as the Jacobi identity for a Lie algebra. It has 15 independent contributions which can be organised into four categories:
\begin{enumerate}
  \item \underline{Closure of the $F$-action on the mega-space} (6 constraints)\\
    These components arise from the requirement that the action of the generalised structure group on the mega-space closes. They are governed by \eqref{eqn:closureFaction}. Only three of them are non-trivial.
    \begin{enumerate}
      \item $\epsilon$-dimension -2 implements the Jacobi identity of Lie($F$),
        \begin{equation}\label{eqn:jacLieF}
          3 f_{[\alpha\beta}{}^\epsilon f_{\gamma]\epsilon}{}^\delta = 0\,.
        \end{equation}
      \item $\epsilon$-dimension -2 also requires that $f_{\alpha B}{}^C$ generates this action on the generalised tangent space $T M_n \oplus T^* M_n$,
        \begin{equation}
          2 f_{[\alpha|C}{}^E f_{|\beta]E}{}^D = - f_{\alpha\beta}{}^\epsilon f_{\epsilon C}{}^D \,.
        \end{equation}
      \item $\epsilon$-dimension 0 imposes the condition
        \begin{equation}\label{eqn:cocycle_1}
        f_{\alpha \beta}{}^\epsilon f_\epsilon{}^{\gamma \delta} 
        - 4 f_{[\alpha| \epsilon}{}^{[\gamma|} f_{|\beta]}{}^{\epsilon |\delta]}
        = 0~.
        \end{equation}
    A simple solution is 
    $f_{\alpha}{}^{\beta \gamma} = -2 \,\rr^{\delta [\beta} f_{\delta \alpha}{}^{\gamma]}$
    for antisymmetric $\rr^{\alpha\beta}$. This condition (and the existence of
    possibly other non-trivial solutions)
    can be understood using the language of Lie algebra cohomology (see e.g. appendix
    A of \cite{Hassler:2016srl}, whose conventions we follow here). One introduces
    one-forms $\theta^\alpha$ (essentially the left-invariant one-forms on $F$) as well as
    scalar 0-forms $e_A$ valued in some representation -- here we are concerned with
    $\Lambda^2 {\rm Lie}(F)$, so we denote $e_{\alpha\beta} = t_\alpha \wedge t_\beta$,
    where the wedge product merely emphasizes that the result is antisymmetrized in
    $\alpha\beta$.
    Then $f_{\alpha}{}^{\beta \gamma}$ is a one-form $\varphi$
    valued in $\Lambda^2  {\rm Lie}(F)$, i.e.
        \begin{equation}
        \varphi = \frac12 \theta^\gamma\, f_\gamma{}^{\alpha\beta} \, t_\alpha\wedge t_\beta ~.
        \end{equation}
    The condition \eqref{eqn:cocycle_1} amounts to closure of this one-form, where we take the
    exterior derivative to obey
        \begin{equation}
          \dd \theta^\alpha = -\frac12 \theta^\beta \wedge \, \theta^\gamma f_{\beta\gamma}{}^{\alpha} 
            \qquad \text{and} \qquad
          \dd t_\alpha = \theta^\beta f_{\beta \alpha}{}^\gamma t_\gamma \,.
        \end{equation}
        Nilpotence of the exterior derivative is guaranteed by \eqref{eqn:jacLieF}.
        Then one indeed finds
        \begin{equation}\label{eqn:cocycle_2}
          \dd \varphi = \frac14 \theta^\alpha \wedge \theta^\beta
          \left( -f_{\alpha\beta}{}^\epsilon f_\epsilon{}^{\gamma\delta}{} + 4 f_{\alpha\epsilon}{}^\gamma f_{\beta}{}^{\epsilon\delta} \right) 
          t_\gamma \wedge t_\delta = 0 
        \end{equation}
        so $\varphi$ must be closed (i.e. a cocycle).
        An immediate solution is to take $\varphi$ to be exact
        (i.e. a coboundary),
        \begin{equation}\label{eqn:varphifromr}
          \varphi = \dd \rr~, \qquad
          \rr = \frac12 r^{\alpha\beta} t_\alpha \wedge t_\beta \, \quad \implies \quad
          f_{\alpha}{}^{\beta \gamma} = -2 \,\rr^{\delta [\beta} f_{\delta \alpha}{}^{\gamma]}~.
        \end{equation}
        But in general, there may also be cocycles that are not coboundaries and therefore cannot be written as $\varphi = \dd \rr$. These are governed by the Lie algebra 
        cohomology $H^1({\rm Lie}(F), \Lambda^2 {\rm Lie}(F))$.
    \end{enumerate}

    \item \underline{Invariance under the $F$-action} (4 constraints)\\
      Next, we find four constraints describing how the components in the 
last column of \eqref{eqn:decompfhABC} transform under the infinitesimal 
action of the generalised structure group $F$.
      \begin{enumerate}
        \item $\epsilon$-dimension $-1$ requires that the torsion 
$f_{ABC}$ is invariant, namely
          \begin{equation}
            3 f_{\alpha [B}{}^E f_{CD]E} = 0\,.
          \end{equation}
          An alternative way to write this equation is by using 
$\nabla_\alpha$:
          \begin{equation}
            \nabla_\alpha f_{BCD} = 0\,.
          \end{equation}
        \item $\epsilon$-dimension $0$ gives rise to
          \begin{equation}
            \nabla_\alpha f_{BCDE} - f_{\alpha}{}^{\beta\gamma} f_{\beta BC} f_{\gamma DE} = 0\,.
          \end{equation}
          By using \eqref{eqn:varphifromr}, we can further simplify this equation to
          \begin{equation}\label{eqn:invarf4}
            \nabla_\alpha \left( f_{BCDE} + \rr_{BCDE} \right) = 0
          \end{equation}
          with
          \begin{equation}
            \rr_{ABCD} := \rr^{\alpha\beta} f_{\alpha AB} f_{\beta CD} \,.
          \end{equation}
          This is quite interesting, because it tells us that $f_{ABCD}$ does not need to be fully invariant under $F$-action. Instead the invariant quantity is rather the sum of $f_{ABCD}$ and the newly introduced $\rr_{ABCD}$. Note that $\rr_{ABCD}$ is antisymmetric under the exchange (AB)$\leftrightarrow$(CD). This implies that the generalised Riemann tensor $\RR_{ABCD}$ defined by \eqref{eqn:ftoRR} is always invariant, as
          \begin{equation}
            \nabla_\alpha \RR_{ABCD} = 0
          \end{equation}
          has to hold.
        \item $\epsilon$-dimension $1$ requires
          \begin{equation}
            \nabla_\alpha f_{ABCDE} = 0
          \end{equation}
          with
          \begin{equation}
            f_{ABCDE}: = f_A{}^{\alpha\beta} f_{\alpha BC} f_{\beta DE} \,.
          \end{equation}
        \item $\epsilon$-dimension $2$ relates $f_{\alpha}{}^{\beta\gamma}$ with $f^{\alpha\beta\gamma}$,
          \begin{equation}
            3 f_\alpha{}^{\epsilon[\beta} f_{\epsilon}{}^{\gamma\delta]} + 3 f_{\alpha\epsilon}{}^{[\beta} f^{\gamma\delta]\epsilon} = 0\,.
          \end{equation}
          With \eqref{eqn:varphifromr}, this expression simplifies to
          \begin{equation}\label{eqn:invarf6}
            \nabla_\alpha ( f_{BCDEFG} + \rr_{BCDEFG} ) = 0
          \end{equation}
          with
          \begin{equation}
            \begin{aligned}
              \rr_{ABCDEF} &=: \rr^{\alpha\beta\gamma} f_{\alpha AB} f_{\beta CD} f_{\gamma EF}\,,\\
              f_{ABCDEF}   &=: f^{\alpha\beta\gamma} f_{\alpha AB} f_{\beta CD} f_{\gamma EF}\,,
            \end{aligned}
          \end{equation}
          and the classical Yang-Baxter tensor
          \begin{equation}
            \rr^{\alpha\beta\gamma} = \rr^{\delta\alpha} \rr^{\beta\epsilon} f_{\delta\epsilon}{}^\gamma + \rr^{\delta\gamma} f_{\delta}{}^{\alpha\beta}\,.
          \end{equation}
          We use this terminology because $\rr^{\alpha\beta\gamma}$ 
 describes the left hand 
side of the classical Yang-Baxter equation. It is manifestly antisymmetric with respect to its first two indices, $\alpha$ and $\beta$. After substituting $f_\alpha{}^{\beta\gamma} = - 2 \rr^{\delta[\beta} f_{\delta\alpha}{}^{\gamma]}$ one can show that it is actually totally antisymmetric.
      \end{enumerate}
    \item\label{item:bianchi} \underline{Bianchi identities} (3 constraints)\\
      Next, we find Bianchi identities for the three curvatures $f_{ABCD}$, $f_{ABCDE}$ and $f_{ABCDEF}$:
      \begin{enumerate}
        \item $\epsilon$-dimension 0 gives the Bianchi identity for $f_{ABCD}$ in \eqref{eqn:algBIgenRiem}, which we already discussed in section~\ref{sec:torsionandcurvature}.
        \item $\epsilon$-dimension 1 implies
        \begin{equation}\label{eqn:BIfABCDE}
          f_{[ABC]DE} = f_{[AB|}{}^F f_{F|C]DE}\,.
        \end{equation}
        To better understand the implications of this equation, we decompose
        \begin{equation}\label{eqn:decompfhABCDE}
          f_{ABCDE} = \ydiagram{2,1,1,1} \oplus \ydiagram{2,2,1} \oplus \ydiagram{3,1,1}\,.
        \end{equation}
        The antisymmetrisation on the right hand side of \eqref{eqn:BIfABCDE} projects out the last contribution. If we also take into account the left hand side, we find
        \begin{equation}
          \ydiagram{2,1,1,1} \oplus \ydiagram{2,2,1} = \ydiagram{2,1,1,1} \oplus \ydiagram{2,2,1} \oplus \overbrace{\ydiagram{1,1,1,1,1}}^{=0}\,,
        \end{equation}
        showing that the totally antisymmetric contribution from the left hand side has to vanish. As result, we obtain the first quadratic constraint, namely
        \begin{equation}
          f_{[AB|}{}^F f_{F|CDE]} = 0 \,.
        \end{equation}
        \item $\epsilon$-dimension 2 results finally in
          \begin{equation}
            \begin{aligned}
              f_{ABCDEF} = - & \frac12 f_{AB}{}^G f_{GCDEF} + f_{G[A|CD} f^G{}_{|B]EF} + f_{G[A|CD} f^G{}_{|B]EF} + \\ 
                             & 2  f_{AB[C}{}^Gf_{|G|D]EF} - (CD)\leftrightarrow (EF)\,.
            \end{aligned}
          \end{equation}
          The right hand side is by definition totally antisymmetric under pairwise exchange of $(AB)$, $(CD)$ and $(EF)$ but the left hand side is not. Therefore we find additional quadratic constraints.
      \end{enumerate}
    \item \underline{Additional quadratic constraints} (2 constraints)\\
      There are two more constraints that do not contain any linear 
contribution. Together with the quadratic constraints we just encountered 
in point~\ref{item:bianchi}, they represent the real challenge in 
identifying admissible curvatures that result in a dressing coset. 
There is not much more we can do about them at the moment, and we therefore
just list them here:
      \begin{enumerate}
        \item $\epsilon$-dimension 3 gives rise to
        \begin{equation}
          -\frac{1}{2}f_{ABIJ}f^B{}_{CDEF}-2 f_{ACD[I}{}^G f_{|G|J]EF} + \asym(IJ,CD,EF) = 0\,,
        \end{equation}
        where $\asym(IJ,CD,EF)$ denotes all permutations of the index pairs $(IJ)$, $(CD)$, $(EF)$ in the given expression weighted by $-1$ for odd permutations.
        \item $\epsilon$-dimension 4 finally results in
        \begin{equation}
      f_{IJCD[K}{}^G f_{|G|L]EF}+\frac{1}{8}f_{ACDKL}f^A{}_{IJEF} + \asym(CD,IJ,EF,KL) = 0\,.
        \end{equation}
      \end{enumerate}
\end{enumerate}
Next, we decompose the second constraint \eqref{eqn:bianchiFhA}. It has six contributions which again can be ordered according to their $\epsilon$-dimension:
\begin{itemize}
  \item $\epsilon$-dimension $-2$ requires
    \begin{equation}
      f_\gamma f^\gamma{}_{\alpha\beta} = 0\,,
    \end{equation}
    which is automatically satisfied for $f_\gamma = f_{\gamma\delta}{}^\delta$.
\item $\epsilon$-dimension $-1$ imposes
    \begin{equation}\label{eqn:fCsinglet}
      \nabla_\alpha f_B = 0\,,
    \end{equation}
    saying that the one-index torsion is a singlet under $F$-action. 
    This is required by theorem~\ref{th:CT}.
  \item $\epsilon$-dimension $0$ gives rise to the two constraints
    \begin{align}\label{eqn:fABfromrestconst}
      f_{AB} &= -f_{AB}{}^C f_C -f_\gamma f^{\gamma}{}_{AB} + f_\gamma{}^{\gamma\delta} f_{\delta AB}\\
      \text{and} \qquad f_{\gamma\alpha}{}^\beta f^\gamma &= -f_\gamma f_\alpha{}^{\beta \gamma}\,.
    \end{align}
    The first equation is used to fix $f_{AB}$. (Recall that 
we did exactly the same in \eqref{eqn:fABfromrest} for the more 
general case of non-constant $f$'s.) In the second equation, we contract the free index $\beta$ with $f_{\beta BC}$ to obtain
    \begin{equation}
      \nabla_\alpha (f_{BC}+\rr_{BC}) = 0 \quad \text{where}\quad \rr_{AB}:= r^{\alpha\beta} f_{\beta\alpha}{}^\gamma f_{\gamma AB} = 2\,\rr^C{}_{ABC}\,.
    \end{equation}
    Satisfying this equation requires that the combination $f_{AB} + \rr_{AB}$
    to be a singlet. This result is not very surprising, because we observed the same phenomena already in \eqref{eqn:invarf4} and \eqref{eqn:invarf6}.
\end{itemize}
There are two further constraints that do not admit such nice interpretations. In general, we note that the higher the $\epsilon$-dimensions becomes, the more complicated it gets. The only advantage here is that both are linear in the components of $f_{\Ah}$. Therefore, they can be always solved easily:
\begin{itemize}
  \item $\epsilon$-dimension $1$ results in
    \begin{equation}
      f^A f_{ABCD}-f_\alpha f_B{}^{\alpha\beta }f_{\beta CD} = 0\,,
    \end{equation}
    while
    \item $\epsilon$-dimension $2$ requires
    \begin{equation}
      f^C f_{CABDE}+f^\gamma f_\gamma{}^{\alpha \beta}f_{\alpha AB}f_{\beta DE}+f_\gamma f^{\gamma \alpha \beta}f_{\alpha AB}f_{\beta DE}=0\,.
    \end{equation}
\end{itemize}

Concluding, the most problematic are the quadratic constraints 
that originate from \eqref{eqn:bianchiFhABC}. There is currently no 
obvious way to find general solutions. Instead one has to treat them on 
a case by case basis.

\subsection{Higher derivative curvature}\label{sec:higherderiv}

In subsection~\ref{sec:truncgenRicci}, one might have gained the 
impression that the components $f_A{}^{\beta\gamma}$ and 
$f^{\alpha\beta\gamma}$ are irrelevant for the action or its equations of 
motion. In fact, this is only true at the leading two-derivative level. 
If we want to consider higher-derivative corrections, there is no obvious 
reason why they should not contribute.

To make this point clear, we first compute
\begin{equation}
  \begin{aligned}
    f_A{}^{\beta\gamma} = &D_A \rho^{\beta\gamma} 
      + D_A \Omega^{[\beta}{}_D \Omega^{\gamma] D} 
      - 2 D_D \Omega^{[\beta}{}_A \Omega^{\gamma] D} 
      - \Omega^\delta{}_A f_\delta{}^{\beta\gamma}
      + F_{ADE} \Omega^{\beta D} \Omega^{\gamma E} - \\ 
      &\Omega^\delta{}_A f_{\delta\epsilon}{}^{[\beta} \Omega^{\gamma]F} \Omega^{\epsilon}{}_F 
      - 2\Omega^\delta{}_A f_{\delta\epsilon}{}^{[\beta} \rho^{\gamma]\epsilon} 
      + \cancel{2 f_{\delta A}{}^{[\beta} \rho^{\gamma]\delta}}
      + \cancel{ f_{\delta A}{}^{[\beta} \Omega^{\gamma] E} \Omega^\delta{}_E } \,.
  \end{aligned}
\end{equation}
As before, we rewrite this quantity in terms of indices on 
$T M \oplus T^* M$ only, as
\begin{equation}
\begin{aligned}
  f_{ABCDE} &= \frac{1}{2} D_A r_{BCDE}  +2 \Omega_{A}{}^F{}_{[B}r_{C]FDE}-2 \Omega_{A}{}^F{}_{[B}f_{C]FDE}-D_F \Omega_{ABC} \Omega^F{}_{DE}\\
  &+\frac{1}{2}D_A \Omega_{FED}\Omega^F{}_{BC}-\Omega_{AI[E}\Omega^{FI}{}_{D]}\Omega_{FBC}-\frac{1}{2}F_{FAG}\Omega^F{}_{BC}\Omega^G{}_{DE}\\
  &-(BC)\leftrightarrow (DE)
\end{aligned}
\end{equation}
There is one more component left, $f^{\alpha\beta\gamma}$, which we have 
to compute. In analogy with $f^\alpha$, it is not independent of the ones 
we computed already. Rather, it arises from the Bianchi identity
\begin{equation}
  2 \nabla^{[\alpha}f^{\beta]}{}_{CD}+2\nabla_{[C}f_{D]}{}^{\alpha\beta}-f^{\alpha \beta \Eh}f_{CD\Eh}+2 f^{\Eh [\alpha}{}_{[C}f^{\beta]}{}_{D]\Eh}=0
\end{equation}
which eventually gives rise to
\begin{equation}
  \begin{aligned}
    f_{ABCDEF} = & r_A{}^G{}_{EF}r_{GBCD}+2r_{G[D|EF|}f_{|AB|C]}{}^G-\frac{1}{2} D_G r_{EFCD}\Omega^G{}_{AB}- f_{G[D|EF|}\Omega_{|I|C]}{}^G \Omega^I{}_{AB}\\
    &+r_{G[D|EF|}\Omega^I{}_{C]}{}^G \Omega_{IAB}+\frac{1}{4}\Omega_{IG[D}\Omega^I{}_{|EF|}\Omega_{|K|C]}{}^G \Omega^K{}_{AB}+\frac{1}{6}F^{GHI}\Omega_{HEF}\Omega_{IAB}\Omega_{GCD}\\
    &+\frac{1}{2} D_I \Omega_{GEF}\Omega^G{}_{AB}\Omega^I{}_{CD} + \asym(AB,CD,EF)
  \end{aligned}
\end{equation}

To count the number of derivatives in each component of $f_{\Ah\Bh\Ch}$, 
we define that either $D_A$ or $\Omega_{ABC}$ count as one, because they 
both contribute equally to the covariant derivative $\nabla_A$ defined 
in \eqref{eqn:defnabla}. Hence, from the \PS{} construction, we find the following independent quantities:
\begin{equation}
  \begin{tabular}{l|c|c|c}
    derivatives & 1 & 2 & 3 \\
    \hline
    quantity & $f_A$\,, $f_{ABC}$ & $f_{ABCD}$ & $f_{ABCDE}$ \\
  \end{tabular}
\end{equation}

At the two-derivative level, we have seen that the action and its equations 
of motion incorporate only the one-derivative torsions $f_A=T_A$ and 
$f_{ABC}=-T_{ABC}$, and the projectors of the two derivative curvature 
$f_{ABCD}$. Starting with the leading $\alpha'$-corrections at four 
derivatives, the situation becomes more complicated because the form of the 
action \cite{Marques:2015vua} is only fixed up to field redefinitions. 
If it is possible to find a field basis in which the generalised structure group $F$ still acts by $\nabla_\alpha$, then the action and its field equations 
should also incorporate $f_{ABCDE}$. Moreover, the singlet condition that 
is required by theorem~\ref{th:CT} has to be imposed at each order of 
$\alpha'$ separately. This will likely result in more constraints in 
addition to \eqref{eqn:constraints2deriv}. Thus, on a qualitative level, we 
note that consistent truncations with higher derivative corrections should 
require that more and more components of the mega-space generalised 
fluxes $f_{\Ah\Bh\Ch}$ are constant. Taking the Bianchi identities into 
account, this might even at some order prohibit any non-constant 
contributions. In this case, there would be a one-to-one correspondence 
between consistent truncations, \`a la theorem~\ref{th:CT}, and generalised cosets. We leave the exploration of this idea to future work.

\subsection{Gauge transformations}\label{sec:gaugetransformations}
We observed in section~\ref{sec:truncgenRicci} that, by what appeared to be 
a miracle, all naked connections in the rewriting of the action and its 
equations of motion vanish and just the various $f$'s remain. Of course 
this is not actually a miracle, but rather the consequence of a 
gauge symmetry. To see how this symmetry emerges, we define its 
infinitesimal action on the mega-space by
\begin{equation}
  \delta_{\lambdah} \Eh_{\Ah\Bh} = (\LL_{\lambdah} \Eh_{\Ah}{}^{\Ih} ) \Eh_{\Bh\Ih} \,.
\end{equation}
A short calculation, using the definition of the generalised Lie derivative \eqref{eqn:genLie}, gives rise to
\begin{equation}\label{eqn:deltaEh}
  \delta_{\lambdah} \Eh_{\Ah\Bh} = - 2 \Dh_{[\Ah} \lambdah_{\Bh]} + \lambdah^{\Ch} \fh_{\Ch\Ah\Bh}\,.
\end{equation}
Essential for our purpose is that all transformations generated by $\lambdah^{\Ah}$ do not change the lower-triangular form of $\Eh_{\Ah}{}^{\Ih}$ given in \eqref{eqn:mega-vielbein}. To see why, we first note that
\begin{equation}
  \lambda^{\Ah} = \Mt_{\Bh}{}^{\Ah} \lambdah^{\Bh}
\end{equation}
should only depend of the internal coordinates $y$ but not on the auxiliary coordinates $z$. In this case, \eqref{eqn:deltaEh} implies
\begin{equation}\label{eqn:deltaE}
  \delta_\lambda E_{\Ah\Bh} = - 2 \nabla'{}_{[\Ah} \lambda_{\Bh]} + \lambda^{\Ch} f_{\Ch\Ah\Bh}\,.
\end{equation}
On the other hand, we can also compute $\delta_\lambda E_{\Ah\Bh}$ directly from the variation of \eqref{eqn:mega-vielbein}, giving rise to
\begin{equation}
  \delta_\lambda E_{\Ah\Bh} = \begin{pmatrix}
    0 & 0 & 0 \\
    0 & \delta_\lambda E_{AB} & \quad - \delta_\lambda \Omega^\beta{}_A - \Omega^\beta{}_C \delta_\lambda E^C{}_B  \quad\\
    0 & \quad \delta_\lambda \Omega^\alpha{}_B + \Omega^\alpha{}_C \delta_\lambda E^C{}_B \quad & \delta_\lambda \rho^{\alpha\beta} + 2 \Omega^{[\alpha}{}_C \delta_\lambda \Omega^{\beta]C} + \Omega^\alpha{}_C \delta_\lambda E^{CD} \Omega^\beta{}_D 
  \end{pmatrix}\,.
\end{equation}
We see two important things here: First, the transformation of 
$E_A{}^I$, $\Omega^\alpha{}_B$ and $\rho^{\alpha\beta}$ can easily be extracted 
from this result, and second, 
$\delta E_{\alpha\Bh} = \delta E_{\Bh\alpha} = 0$. The latter is not 
automatically guaranteed by \eqref{eqn:deltaE}. It requires the restriction
\begin{equation}\label{eqn:lambdacomponents}
  \lambda_{\Ah} = \begin{pmatrix} \lambda^\alpha & \lambda_A & 0 \end{pmatrix}\,.
\end{equation}
But the second parameter $\lambda_A$ only mediates a generalised diffeomorphism on the physical space $M_d$. For the following discussion, we therefore concern ourselves only with $\lambda^\alpha$, which mediates gauge transformations, and compute
\begin{align}
  \delta_\lambda E_{AB} &= \lambda^\gamma f_{\gamma AB} \,, \\
  \delta_\lambda E^{\alpha}{}_B &= D_B \lambda^\alpha + \Omega^\gamma{}_B f_{\gamma\delta}{}^\alpha \lambda^\delta - \cancel{\lambda^\gamma f_{\gamma B}{}^{\alpha}} \,, \quad \text{and} \\
  \delta_\lambda E^{\alpha\beta} &= 2 D_C \lambda^{[\alpha} \Omega^{\beta]C} - 2 \rho^{\gamma[\alpha} f_{\gamma\delta}{}^{\beta]} \lambda^\delta - \Omega^{\gamma I} \Omega^{[\alpha}{}_I f_{\gamma\delta}{}^{\beta]} \lambda^\delta + \lambda^\gamma f_{\gamma}{}^{\alpha\beta}\,,
\end{align}
eventually extracting
\begin{align}
  \delta_\lambda \Omega^\alpha{}_B &= D_B 
\lambda^\alpha-\Omega^\alpha{}_C \lambda^C{}_B +
\Omega^\gamma{}_B f_{\gamma \delta}{}^\alpha \lambda^\delta\,,\\
  \delta_\lambda \rho^{\alpha\beta} &=4 D_C \lambda^{[\alpha}\Omega^{\beta]C}-2\rho^{\gamma[\alpha}f_{\gamma\delta}{}^{\beta]}\lambda^\delta -3\Omega^{[\alpha}{}_{E}f_{\gamma\delta}{}^{\beta]}\Omega^{\gamma E} \lambda^{\delta}+3\Omega^{[\alpha}{}_{C}\Omega^{\beta]}_{D}\lambda^{D C}+\lambda^\gamma f_\gamma{}^{\alpha \beta}\,.
\end{align}
As in earlier calculations, it is useful to write the results just with 
indices for the generalised tangent space on the internal manifold 
$T M \oplus T^* M$. To do so, we introduce
\begin{equation}
  \lambda_{AB} := \lambda^\gamma f_{\gamma AB}
\end{equation}
and obtain
\begin{align}
  \delta_\lambda E_{AB} &= \lambda_{AB}\,,\\
  \delta_\lambda \Omega_{ABC} &= D_A\lambda_{BC}-\lambda_A{}^D \Omega_{DBC}-2\lambda_{D[C}\Omega_{|A|B]}{}^D \,, \\
  \delta_\lambda r_{ABCD} &= 2
  D_{E}\lambda_{AB}\Omega^E{}_{CD}-2D_{E}\lambda_{CD}\Omega^E{}_{AB}+ 2 \lambda^E{}_{[D}r_{C]EAB}-2\lambda^E{}_{[B} r_{A]ECD} \nonumber\\
    & + 3 \Omega_{FAB}\lambda^E{}_{[D}\Omega^F{}_{C]E}-3
    \lambda_{E[B}\Omega^F{}_{A]E}\Omega_{FCD}+3\Omega_{EAB}\Omega_{FCD}\lambda^{EF} \nonumber\\
    & + 2 \lambda_{[A}{}^G f_{|G|B]CD}+2\lambda_{[C}{}^G f_{|ABG|D]}\,.
\end{align}

Here, we recognise the double Lorentz transformation rules for the frame and the spin connection $\Omega_{ABC}$. Additionally, there is a new
transformation rule for the \PS{} field $r_{ABCD}$. To understand how it arises, we look at the transformation of the generalised fluxes $\fh_{\Ah\Bh\Ch}$ on the mega-space. Under generalised diffeomorphisms they transform as scalars,
\begin{equation}
  \delta_{\lambdah} \fh_{\Ah\Bh\Ch} = \lambdah^{\Dh} \Dh_{\Dh} \fh_{\Ah\Bh\Ch}\,,
\end{equation}
or, equally,
\begin{equation}\label{eqn:deltalambdafAhBhCh}
  \delta_\lambda f_{\Ah\Bh\Ch} = \lambda^\delta \nabla'{}_{\delta} f_{\Ah\Bh\Ch}\,.
\end{equation}
A similar equation,
\begin{equation}
  \delta_\lambda f_{\Ah} = \lambda^\beta \nabla'{}_{\beta} f_{\Ah}\,,
\end{equation}
holds for the one-index generalised flux $\fh_{\Ah}$, too. These two 
equations become even more useful if we exchange the derivative 
$\nabla'{}_\alpha$ for $\nabla_\alpha$, which captures the covariant 
action of the generalised structure group on $T^M \oplus T^*M$. 
Various quantities we already encountered do not transform covariantly. 
Examples are the connections $\Omega_{ABC}$ and $r_{ABCD}$. To quantify 
the deviation from a covariant transformation, it is convenient to 
introduce the operator
\begin{equation}
  \Delta_\lambda = \delta_\lambda - \lambda^\alpha \nabla_\alpha \,.
\end{equation}
Any quantity it annihilates is covariant. In general, there are two perspectives one can take on the \PS{} construction. We started by looking at the extended, mega-space with $\fh_{\Ah\Bh\Ch}$ and $\Eh_{\Ah}{}^{\Ih}$. On the other hand, it also has to be possible to extract the same information just from the internal space $M$. Indeed this is when we consider the following data on $M$:
\begin{enumerate}
  \item A generalised structure group $F$ which acts faithfully on the generalised tangent space $T M \oplus T^* M$. This will fix $f_{\alpha B C}$, $f_{\alpha\beta}{}^\gamma$ and set $f_{\alpha B}{}^\gamma = 0$.
  \item All dynamic contributions to the generalised fluxes on the 
mega-space. They are listed in the last columns of \eqref{eqn:decompfhABC} 
and \eqref{eqn:complistfAh} which are completely encoded in the torsions 
and curvatures
    \begin{center} 
      $f_A$, $f_{ABC}$, $f_{ABCD}$ \quad and \quad $f_{ABCDE}$\,.
    \end{center}
  \item Everything else is fixed by the Bianchi identity on the mega-space.
\end{enumerate}
Taking into account \eqref{eqn:deltalambdafAhBhCh}, we find that all torsions and curvatures mentioned transform covariantly, except for
\begin{equation}
  \Delta_\lambda f_{ABCD} = \lambda^\alpha f_{\alpha}{}^{\beta\gamma} f_{\beta AB} f_{\gamma CD}\,.
\end{equation}

This is the power of the \PS{} formalism. It provides a systematic method 
for finding covariant tensors. Moreover, we know that both the 
generalised Ricci scalar $\RR$ and the generalised Ricci tensor 
$\RR_{AB}$ transform covariantly under double Lorentz transformations. 
Hence, when we rewrite them in terms of $f$'s in 
section~\ref{sec:truncgenRicci}, they must remain covariant. Therefore, 
all the non-covariant naked connections $\Omega_{ABC}$ must in the end
be formed from covariant $f$'s or covariant derivatives of them. We 
also see that already at the two derivative level the antisymmetric 
part (with respect to $(AB) \leftrightarrow (CD)$) of $f_{ABCD}$ is not covariant. Thus, it cannot appear in the two derivative action or its field equations.

\section{Construction of the mega-generalised frame}\label{sec:vielbeins}
Currently all known generalised T-dualities can be captured by 
dressing cosets. In section~\ref{sec:truncgenRicci}, we concluded that these 
cosets result in consistent truncations where the \PS{} construction has 
constant generalised fluxes $\fh_{\Ah\Bh\Ch}$ on the mega-space. 
The objective now is to show how we can construct the corresponding 
mega-generalised frames that make the truncation ansatz discussed in 
section~\ref{sec:trunctheory} fully explicit. As we have seen in 
section~\ref{sec:gSSandPLTD}, these frames are not only very valuable for 
consistent truncations, but they also are an important tool for studying 
generalised T-dualities on the worldsheet, and for constructing 
the canonical transformations of the underlying \s-models.

\subsection{Generalised frames on group manifolds}\label{sec:genframegroup}
Our starting point is the constant generalised fluxes on the mega-space,
 and we are looking for a generalised frame $\Eh_{\Ah}{}^{\Ih}$ that 
satisfies
\begin{equation}\label{eqn:fhconst}
  \fh_{\Ah\Bh\Ch} = 
\Dh_{[\Ah} \Eh_{\Bh}{}^{\Ih} \Eh_{\Ch]\Ih} = \text{const.}
\end{equation}
As outlined in section~\ref{sec:Jac}, the corresponding Bianchi identity 
becomes the Jacobi identity \eqref{eqn:bianchiFhABC} of a Lie algebra. 
We denote it as Lie($\DD$) and interpret the generalised 
fluxes $\fh_{\Ah\Bh\Ch}$ as its structure coefficients. An important 
consequence is that the adjoint action of any element of $\DD$ will 
leave $\fh_{ABC}$ invariant. Therefore, we can identify 
$\fh_{\Ah\Bh\Ch} = f_{\Ah\Bh\Ch}$. It is known that one can 
construct $\Eh_{\Ah}{}^{\Ih}$ systematically 
\cite{Hassler:2017yza,Demulder:2018lmj,Hassler:2019wvn,Borsato:2021vfy}, 
based on the following data:
\begin{enumerate}
  \item\label{prop1} A doubled Lie group $\DD$, which is generated by the generators $t_{\Ah}$ with the structure coefficients
    \begin{equation}
      [ t_{\Ah}, t_{\Bh} ] = f_{\Ah\Bh}{}^{\Ch} t_{\Ch}\,.
    \end{equation}
    In the following, we label them by $t_{\Ah} = \begin{pmatrix} t_{\ah} & t^{\ah} \end{pmatrix}$, where $\Ah = 1, \dots, 2 \Dh$ and $\ah=1, \dots, \Dh$.
  \item A non-degenerate, pairing, $\langle t_{\Ah}, t_{\Bh} \rangle = \eta_{\Ah\Bh} = \begin{pmatrix} 0 & \delta_{\hat a}{}^{\hat b} \\
    \delta^{\hat a}{}_{\hat b} & 0 \end{pmatrix}$, that is invariant under the adjoint action of $\DD$.
  \item\label{prop3} A maximally isotropic subgroup $H \subset \DD$, generated by $t^{\hat a}$, with $\langle t^{\ah} , t^{\bh} \rangle = 0$.
\end{enumerate}
Explicitly, the generalised frame is defined on the coset $M = H \backslash \DD$ in terms of
\begin{equation}\label{eqn:megagenframe}
  \Eh_{\Ah}{}^{\Ih} = M_{\Ah}{}^{\Bh} \begin{pmatrix}
    \vh_{\bh}{}^{\ih} & \vh_{\bh}{}^{\jh} B_{\jh\ih} \\
    0 & v^{\bh}{}_{\ih}
  \end{pmatrix}\,.
\end{equation}
In studying its properties, it is convenient to use the differential forms $v^{\ah} = v^{\ah}{}_{\ih} \dd x^{\ih}$, $A_{\ah} = A_{\ah\ih} \dd x^{\ih}$, $B=\frac12 B_{\ih\jh} \dd x^{\ih}\wedge x^{\jh}$ which are defined by
\begin{align}
\label{eqn:megagenframe.dm}
\dd m m^{-1} & = t_{\ah} v^{\ah} + t^{\ah} A_{\ah}\,, \qquad m \in M \\
\label{eqn:megagenframe.B}
  B &= \frac12 v^{\ah} \wedge A_{\ah} + B_{\WZW}\,, \\
\label{eqn:megagenframe.WZW}
  \dd B_{\WZW} = H_{\WZW} &= -\frac1{12} \langle \dd m m^{-1}, [ \dd m m^{-1}, \dd m m^{-1} ] \rangle\,.
\end{align}
In general, $H_{\WZW}$ is closed but not exact. If this is the case, 
$B_{\WZW}$ can only be defined in a local patch and the patches have 
to be connected by appropriate gauge transformations. Moreover, we need 
the adjoint action
\begin{equation}
  m t_{\Ah} m^{-1} = M_{\Ah}{}^{\Bh} t_{\Bh}\,,
\end{equation}
and the dual vector field $\vh_{\ah}{}^{\ih}$, 
defined by $\vh_{\ah}{}^{\ih} v^{\bh}{}_{\ih} = \delta_{\ah}{}^{\bh}$, 
to complete the list of ingredients that enter \eqref{eqn:megagenframe}.

At this point, we finally fully review how Poisson-Lie T-duality is 
realised by generalised Scherk-Schwarz reductions. Looking at the 
diagram in \eqref{eqn:dualitiesdiagram}, we want to preserve the 
truncated theory, and, most importantly, with it the generalised 
torsions $T_A$ and $T_{ABC}$. With them the generalised Ricci scalar 
$\RR$ and tensor $\RR_{AB}$ are also preserved, because there is no 
curvature $\RR_{ABCD}$; all that counts for generalised Scherk-Schwarz 
reductions are the torsions. To preserve them, we are looking for 
different generalised frames $E^{(i)}_A{}^I$ which still produce the 
same constant generalised fluxes $F_{ABC}$. We used the
construction presented above (after removing all hats) to achieve this 
goal. The generalised fluxes depend only on the double Lie group $\DD$, but 
not on the choice of the maximally isotropic subalgebra. On the other hand, 
the constructed generalised frame $E_A{}^I$ crucially depends on the 
subalgebra used in the construction. For any maximally isotropic 
subalgebra $H_i$, we obtain a new generalised frame field $E^{(i)}_A{}^I$ 
that still gives rise to the same generalised fluxes $F_{ABC}$. On the 
string worldsheet, the same mechanism was used to define Poisson-Lie 
T-duality and to show that it is a canonical transformation between 
two $\sigma$-models with dual target spaces.

We are not completely done yet, because there is still the one-index 
generalised flux $\fh_{\Ah}$. As it is constant, its Bianchi identity on 
the mega-space simplifies to \eqref{eqn:bianchiFhA}. As a consequence 
the adjoint action of any element in $\DD$ will leave $\fh_{\Ah}$ 
invariant and we identify $\fh_{\Ah} = f_{\Ah}$. In analogy with 
\eqref{eqn:fhconst}, we have now to find a $\dh$ such that
\begin{equation}
  \fh_{\Ah} = 2 \Dh_{\Ah} \dh - \partial_{\Ih} \Eh_{\Ah}{}^{\Ih} = 
\text{const.}
\end{equation}
holds. The component
\begin{equation}\label{eqn:fixedcomp}
  f^{\ah} = f_{\bh}{}^{\bh\ah}
\end{equation}
is automatically constant and from $f_{\ah}$, we obtain the differential equation
\begin{align}
  \dd \db &= \frac12 v^{\ah} \left( f_{\ah} - f_{\ah\bh}{}^{\bh} + \iota_{\vh_{\bh}} A_{\ch} f_{\ah}{}^{\bh\ch} \right) - \frac12 \iota_{\vh_{\ah}} \dd v^{\ah} \nonumber \\ 
  &= \frac12 \left( v^{\ah} f_{\ah} + A_{\ah} f^{\ah} \right) = \frac12 \langle \dd m m^{-1}, t_{\Ah} \rangle f^{\Ah}
\label{eqn:ddilaton}
\end{align}
with
\begin{equation}\label{eqn:dconstfAh}
  \dh = \db - \frac12 \log \det v
\end{equation}
that fixes $\dh$ up to a constant. The integrability condition for $\dd^2 \db = 0$ for this equation follows immediately from the Bianchi identity because
\begin{equation}
  \dd^2 \db = \frac14 \langle [\dd m m^{-1}, \dd m m^{-1}], t_{\Ah} \rangle f^{\Ah} = \frac14 v^{\Ah} \wedge v^{\Bh} \, f_{\Ah\Bh}{}^{\Ch} f_{\Ch} = 0\,,
\end{equation}
where $v^{\Ah}$ denotes $\dd m m^{-1} = t_{\Ah} v^{\Ah}$\,. This observation has been already made in \cite{Borsato:2021vfy}. Also one should note that according to \eqref{eqn:fixedcomp} the generalised fluxes $\fh_{\Ah}$ and $\fh_{\Ah\Bh\Ch}$ are not completely independent. One half of the former is completely fixed by the latter. It is not possible to break this connection in supergravity. But there is also the framework of generalised supergravity \cite{ Arutyunov:2015mqj}, where this additional constraint besides the Bianchi identities is not required \cite{Borsato:2021vfy}.

\subsubsection*{$H$-shift of the coset representative and $B$-field gauge transformations}
The coset representative we used in the last section is defined only up to 
the action of $H$ from the left. There, one might ask what happens to the 
generalised frame if we shift $m \rightarrow m' = h m$, with $h\in H$. 
The adjoint action in \eqref{eqn:megagenframe} transforms as
\begin{align}
M'_{\Ah}{}^{\Bh} &= M_{\Ah}{}^{\Ch} \Lambda_{\Ch}{}^{\Bh}~, \qquad 
h T_{\Ah} h^{-1} =: \Lambda_{\Ah}{}^{\Bh} T_{\Bh}~.
\end{align}
The fact that $H$ is an isotropic subgroup guarantees that $\Lambda^{\ah \bh}$ vanishes.
To evaluate the remaining quantities in \eqref{eqn:megagenframe}, we compute first
\begin{align}
\dd m' m'^{-1} &= h \,\dd m m^{-1} \,h^{-1} +\dd h h^{-1}~.
\end{align}
Defining $\dd h h^{-1} = \omega_{\ah} T^{\ah}$, one can show that the one forms $v^{\ah}$ and $A_{\ah}$
shift as
\begin{align}
v'^{\ah} = v^{\bh} \Lambda_{\bh}{}^{\ah}~, \qquad
A'_{\ah} = v^{\bh} \Lambda_{\bh \ah} + A_{\bh} \Lambda^{\bh}{}_{\ah} + \omega_{\ah}~.
\end{align}
For the WZW contribution to the $B$-field, we find
\begin{align}
B'_{\rm WZW} \cong  B_{\rm WZW} - \frac{1}{2} v^{\ah} \Lambda_{\ah}{}^{\bh} \omega_{\bh}
\end{align}
where $\cong$ denotes equality up to an undetermined exact term. This exact term cannot be
determined explicitly: the ansatz \eqref{eqn:megagenframe} involves only a locally defined
$B_{\rm WZW}$ via \eqref{eqn:megagenframe.WZW}. This leads to
\begin{align}
B' &\cong B + \frac{1}{2} v^{\ah} \wedge v^{\bh} \Lambda_{\ah}{}^{\ch} \Lambda_{\bh \ch}~.
\end{align}
The combination $\Lambda_{\ah}{}^{\ch} \Lambda_{\bh \ch}$ is antisymmetric in $\ah \bh$
from the orthogonality condition on $\Lambda$.
The difference in these two $B$-fields is not closed, but this is not crucial, because
the field denoted $B$ here is not \emph{precisely} the physical $B$-field; the latter is
encoded in the generalized metric \eqref{eqn:genmetric} and $M_A{}^B$ contributes 
non-trivially. When the shift in $M$ to $M'$ is accounted for, one finds indeed that
it cancels the transformations not only of $B$ but also of $\vh$, leading to
\begin{equation}
\Eh'_{\Ah}{}^{\Ih} 
= M'_{\Ah}{}^{\Bh} \begin{pmatrix}
    \vh'_{\bh}{}^{\ih} & \vh'_{\bh}{}^{\jh} B'_{\jh\ih} \\
    0 & v'^{\bh}{}_{\ih}
  \end{pmatrix} \cong
  M_{\Ah}{}^{\Bh} \begin{pmatrix}
    \vh_{\bh}{}^{\ih} & \vh_{\bh}{}^{\jh} B_{\jh\ih} \\
    0 & v^{\bh}{}_{\ih}
  \end{pmatrix}
  = \Eh_{\Ah}{}^{\Ih} 
\end{equation}
where $\cong$ denotes equality up to an exact shift in the $B$-field.
Hence, changing the coset representative $m$ just amounts to making a 
$B$-field gauge transformation. This is nice, because it implies that 
different possible choices for $m$ are all related to each other,
justifying the identification of the space as a coset.

As one might expect, the generalised dilaton is unchanged (up to a constant) since
it is determined by integrating the expression for $f_{\ah}$.
From \eqref{eqn:ddilaton}, one finds
$\dd \db' = \dd \db + \frac{1}{2} \omega_{\ah} f^{\ah}$
where the shift $\omega_{\ah} f^{\ah}$ is closed. In fact, it is exact since
\begin{align}
\dd \log \det \Lambda_{\ah}{}^{\bh} = f_{\bh}{}^{\bh \ah} \omega_{\ah} = \omega_{\ah} f^{\ah} 
\end{align}
and this relation implies that 
\begin{align}
  \dh' &= \dh + \textrm{const}~.
\end{align}

\subsection{\PS{} form of the mega frame}

Next, we have to check if the generalised frame we just constructed can be 
brought into the \PS{} form \eqref{eqn:mega-vielbein} introduced in 
section~\ref{sec:mega-vielbein}. Because this form is tightly constrained 
by O($\Dh$, $\Dh$), we have only to check the last column, namely
\begin{equation}\label{eqn:columntocheck}
  \Eh_{\Ah}{}_\mu = \Mt_{\Ah}{}^{\Bh} \begin{pmatrix} 0 \\ 0 \\ \vt^\beta{}_\mu \end{pmatrix}\,.
\end{equation}
Considering the insight gained in \cite{Demulder:2019vvh}, we expect that 
this condition can only be satisfied if we consider a dressing coset. 
Therefore, we first decompose the coset representative $m$ as
\begin{equation}
  m = n f \,, \quad \text{with} \quad n \in H \backslash \DD / F \quad \text{and} \quad f \in F\,.
\end{equation}
Now, $n$ is a representative of a double coset, which is called a dressing
coset in the 
context of generalised T-duality \cite{Klimcik:1996np}. In particular, 
$F$ has to be an isotropic subgroup, but not necessarily maximally isotropic. 
The coordinates that we choose to parameterise $n$ are called $y^i$, 
while for $f$, we use $z^\mu$. As before, we adapt all constituents of the 
generalised frame in \eqref{eqn:megagenframe} to this new decomposition, 
starting with
\begin{align}
  M_{\Ah}{}^{\Bh} &= \Mt_{\Ah}{}^{\Ch} \Mb_{\Ch}{}^{\Bh}\,, 
   \qquad\qquad \text{with} &&
  \Mt_{\Ah}{}^{\Bh} t_{\Bh} = f t_{\Ah} f^{-1} \\ &&&
  \Mb_{\Ah}{}^{\Bh} t_{\Bh} = n t_{\Ah} n^{-1} \,.
\end{align}
We also find
\begin{align}\label{eqn:Bfielddressingcoset}
  B &= \frac12 \left( v^{\ah} \wedge A_{\ah} - \langle \dd f f^{-1}, n^{-1} \dd n \rangle \right) + \Bb_{\WZW} \quad \text{with} \\
  \Hb_{WZW} &= \dd \Bb_{\WZW} = - \frac1{12} \langle \dd n n^{-1}, [\dd n n^{-1}, \dd n n^{-1}] \rangle\,,
\end{align}
which allows us to compute
\begin{align}\label{eqn:ivthB}
  \iota_{\vth_{\alpha}} B &= - \langle n t_\alpha n^{-1} , t_{\ah} \rangle v^{\ah} \,, \\
  \iota_{\vth_{\alpha}} \iota_{\vth_{\beta}} B &=  \langle n t_\alpha n^{-1}, t_{\bh} \rangle \langle t^{\bh}, n t_\beta n^{-1} \rangle\,,
\end{align}
by taking into account that $F$ is isotropic. (Recall that 
$\vth_{\alpha}$ is the dual vector field to the Maurer-Cartan form 
$t_\alpha \vt^\alpha{}_{\mu} \dd x^\mu = \dd f f^{-1}$, which 
we introduced in section~\ref{sec:mega-vielbein}.)

Equation \eqref{eqn:columntocheck} can now be equivalently written as
\begin{equation}
  \iota_{\vth_{\beta}} \langle n t_{\Ah} n^{-1}, t^{\ch} \iota_{\vh_{\ch}} B + t_{\ch} v^{\ch} \rangle = \langle t_{\Ah}, t_\beta \rangle\,,
\end{equation}
which is better suited to be checked. 
Using, \eqref{eqn:ivthB}, $v^{\ah} = \langle t^{\ah}, \dd n n^{-1} + 
n \dd f f^{-1} n^{-1} \rangle$, $\iota_{\vth_{\alpha}} \dd f f^{-1} = 
t_{\alpha}$, and $\iota_{\vh_{\ah}} v^{\bh} = \delta_{\ah}^{\bh}$, one can 
easily show that this equation indeed holds. Therefore, just using the 
appropriate $B$-field given in \eqref{eqn:Bfielddressingcoset}, 
the generalised frame field on the mega-space \eqref{eqn:megagenframe} takes 
the \PS{} form.

We also have to check if the generalised dilaton $\dh$, which we fixed in 
\eqref{eqn:dconstfAh}, is compatible with the ansatz \eqref{eqn:fhA} on 
the mega-space. To this end, we first write
\begin{equation}
  v^{\ah}{}_{\ih} = \underbrace{\begin{pmatrix} \eta^{\ah}{}_\beta \quad& \Mb^{\ah}{}_b \end{pmatrix}}_{\displaystyle :=\rmm}
    \begin{pmatrix}
      \vt^\beta{}_\mu & 0 \\
      0 & \vb^b{}_i
    \end{pmatrix}\,,
\end{equation}
with the one-forms $\vb^a$ which are defined by $t_a \vb^a{}_i \dd y^i = 
\dd n n^{-1}$. After plugging this expression into 
\eqref{eqn:dconstfAh}, we obtain
\begin{equation}
  \dh = \db - \frac12 \log \det \rmm(y)  - \frac12 \log \det v(y) - \frac12 \log \det \vt(z)\,.
\end{equation}
Comparing this equation with \eqref{eqn:fhA}, we have to identity
\begin{equation}
  d(y) = \db - \frac12 \log \det \rmm(y)  - \frac12 \log \det v(y) - \frac12 \log \det \widetilde{\rmm}(z)
\end{equation}
with $\widetilde{\rmm}_\alpha{}^\beta = \Mt_\alpha{}^\beta$.
However, this only works if the final result, $d(y)$, does not depend on the auxiliary coordinates $z$. To check if this is indeed the case, we compute
\begin{equation}
  \iota_{\vth_\alpha} \dd d = \iota_{\vth_\alpha} \dd \db - f_{\alpha\beta}{}^\beta = f_\alpha - f_{\alpha\beta}{}^\beta = 0\,.
\end{equation}
Fortunately, this is exactly the constraint we already imposed in 
\eqref{eqn:complistfAh}. Hence, we conclude that also the dilaton is 
of the right \PS{} form.

 This completes the proof of the central result of this paper:
\begin{theorem}\label{th:MAIN}
  For every every dressing coset $H \backslash \DD / F$, where $\DD$ and $H$ satisfy the properties~\ref{prop1}-\ref{prop3} in section~\ref{sec:genframegroup}, and $F$ is an isotropic subgroup of $\DD$, there exists a consistent truncation, governed by theorem~\ref{th:CT}, with generalised structure group $F$.
\end{theorem}

\subsection{$F$-shift and gauge transformations}

We already figured out in section \ref{sec:genframegroup} what happens to 
the coset representative under 
$H$-shifts from the right. To extend this discussion to generalised cosets, we 
now study the infinitesimal $F$-action from the left on the 
representative $n \in H \backslash \DD / F$. To this end, we 
shift $\delta n = n \delta h$ with $\delta h = 
\lambda^\alpha t_\alpha := \lambda$. Under this shift, we first find
\begin{equation}
  \delta M_{\Ah}{}^{\Bh} = \lambda^\alpha \Mt_{\Ah}{}^{\Ch} \fh_{\alpha\Ch}{}^{\Dh} \Mb_{\Dh}{}^{\Bh}
\end{equation}
and after a bit more of calculation
\begin{align}
  \delta (\dd m m^{-1}) &= t_{\ah} \delta v^{\ah} + t^{\ah} \delta A_{\ah} = n ( \dd \lambda +[ \lambda, \dd f f^{-1} ] ) n^{-1}\,, \\
  \delta B_{\WZW} &= - \frac12 \langle \delta (\dd m m^{-1}), \dd n n^{-1} \rangle \,, \quad \text{and} \\
  \delta B &= v^{\ah} \wedge \delta A_{\ah} \,.
\end{align}
We also need the transformation of the one-form $v^{\ah}$ and its dual 
vector fields $\vh_{\ah}$. It is convenient not to treat them separately, 
but instead combine them into
\begin{equation}
  V_{\Ah}{}^{\Ih} = \begin{pmatrix}
    \vh_{\ah}{}^{\ih} & 0 \\
    0 & v^{\ah}{}_{\ih}
  \end{pmatrix}
\end{equation}
and compute 
\begin{equation}
  \delta V_{\Ah\Bh} = \Mb_{\Ah}{}^{\Ch} \delta V_{\Ch}{}^{\Ih} V^{\Dh}{}_{\Ih} \Mb_{\Bh\Dh} = - 2 \iota_{E_{[\Ah}} \delta v^{\ch} \Mb_{\Bh]\ch} \,.
\end{equation}
In the same vein, we introduce
\begin{equation}
  \delta B_{\Ah\Bh} = \Mb_{\Ah}{}^{\ch} \Mb_{\Bh}{}^{\dh} \iota_{\vh_{\dh}} \iota_{\vh_{\ch}} \delta B = -2 \iota_{E_{[\Ah}} \delta A_{\ch} \Mb_{\Bh]}{}^{\ch} \,.
\end{equation}
These two quantities allow us to write the shift of the mega generalised 
frame in the compact form
\begin{equation}
  \delta \Eh_{\Ah\Bh} := \delta \Eh_{\Ah}{}^{\Ih} \Eh_{\Bh\Ih} = \Mt_{\Ah}{}^{\Ch} \Mt_{\Bh}{}^{\Dh} \left( -2 \nabla'{}_{[\Ch} \lambda_{\Dh]} + \lambda_{\Ch\Dh} \right)
\end{equation}
with
\begin{equation}
  \lambda_{\Ah} = \begin{pmatrix} 0 & 0 & \lambda^\alpha \end{pmatrix} \qquad \text{and} \qquad
  \lambda_{\Ah\Bh} = \lambda^\gamma f_{\gamma\Ah\Bh}\,.
\end{equation}
Comparing these equations with \eqref{eqn:deltaE} and \eqref{eqn:lambdacomponents} in section~\ref{sec:gaugetransformations}, we notice that they are exactly same gauge transformations we discussed in section~\ref{sec:gaugetransformations}. Therefore, we know that they leave the constant generalised fluxes $f_{\Ah\Bh\Ch}$ invariant.

\section{Conclusion and outlook}
The results in this paper reveal a new, deep connection between 
consistent truncations and dualities. In particular, we established in 
theorem~\ref{th:MAIN} that generalised cosets, which underlie the most 
general formulation of T-duality currently known (excluding mirror symmetry), 
automatically give rise to a large family of consistent truncations. The 
relation between them is not immediately obvious, and we therefore developed 
a new geometrical approach that makes the generalised structure group of the 
consistent truncation manifest by introducing an auxiliary space.

We showed that the relation represented by the solid arrow in the diagram
\begin{equation}\label{eqn:arrows}
  \begin{tikzpicture}
    \node[name=TD] {generalised T-dualities};
    \node[at={(TD.east)},anchor=west,xshift=5em,name=trunc] {consistent truncations};
    \draw[->] ($(TD.east)+(0,0.1)$) -- ($(trunc.west)+(0,0.1)$);
    \draw[dashed,->] ($(trunc.west)-(0,0.1)$) -- ($(TD.east)-(0,0.1)$);
  \end{tikzpicture}
\end{equation}
holds. This means that all currently known backgrounds which give rise 
to generalised T-dualities also produce consistent truncations. We did not 
yet manage to determine the fate of the other direction, represented by the 
dashed arrow. There are two interesting alternatives that our analysis 
currently hints at:
\begin{enumerate}
  \item Our new approach to generalised cosets suggests that their description 
in generalised geometry automatically leads to curvatures with more than 
two derivatives. This brings into question what happens for consistent 
truncations in (super)gravity beyond the leading, two-derivative level. 
To the best of our knowledge there is currently not much known. The 
results from section~\ref{sec:higherderiv} indicate that there might be 
new constraints that complement those already known from 
theorem~\ref{th:CT}. At the moment it seems that we know more consistent 
truncations than generalised T-dualities. However, these new constraints
could level the field and even in the end result in a one-to-one 
correspondence.
  \item Alternatively, there may also be some new generalised T-dualities, 
waiting to be found. The relation between the former and consistent 
truncations, sketched in diagram \eqref{eqn:dualitiesdiagram}, could
then be a useful tool for searching for new dualities by studying existing 
examples of consistent truncations beyond generalised cosets.
\end{enumerate}
Besides these conceptual questions, there are also important applications for our results. They originate from both sides of~\eqref{eqn:arrows}.
\begin{itemize}
  \item \underline{Generalised T-dualities:} One particularly active 
sub-branch of this field is concerned with the construction and analysis of 
integrable deformations. Although it is still not completely understood why, 
there is a very close relation between generalised T-dualities and 
integrable string theories. The latter are among the primary means of
exploring new concepts in theoretical physics, because they provide a 
superior level of computational control in comparison to models 
which are not integrable. All results we derived here apply mostly to 
bosonic strings or the NS/NS sectors of superstrings. This is sufficient 
for answering conceptual questions, but for concrete applications, 
such as integrable Green-Schwarz strings required for probing 
the AdS/CFT correspondence, the full R/R sector is needed also,
together with supersymmetry. We shall address this problem in a 
forthcoming paper \cite{supergrouppaper}, by extending the results of
the current paper to supergroups, in a supersymmetric version of double 
field theory \cite{Butter:2022gbc} proposed by one of the authors.

  \item \underline{Consistent truncations:} The last years have seen 
significant progress in constructing and understanding 
consistent truncations. They are mostly centered around exceptional 
generalised geometry and exceptional field theory, and have 
applications reaching from the AdS/CFT correspondence to supporting 
or disproving swampland conjectures. Exceptional field theory goes beyond 
the string and incorporates membranes too. At the same time, 
T-duality enhances to U-duality, which takes into account S-duality as well. 
Another advantage is that R/R fluxes are automatically implemented. 
However, one should not think that they arise in the same ways as in the 
Green-Schwarz string discussed above. While, for the latter, 
supersymmetry and its fermionic degrees of freedom result in R/R fluxes, 
in exceptional geometry/field theory U-duality is the driving force. 
In particular, it relates strings and D-branes, which source R/R fluxes. 
One important consequence is that exceptional field theory still requires 
an explicit splitting of the spacetime, which makes it perfectly suited 
for studying consistent truncations in supergravity. Therefore, one should 
also try to extend the results we have presented here for O($n$,$n$) 
generalised geometry to the corresponding exceptional groups $E_{n+1(n+1)}$. 
If possible, and there are no obvious conceptual problems, this would open 
up a route to many new consistent truncations, with a wide range of 
applications. On the other hand, it will also shed new light on extending 
generalised T-duality to U-duality, which has been already initiated 
\cite{Sakatani:2019zrs,Malek:2019xrf} based on the results in double 
field theory.
\end{itemize}

We hope that there will be more insights into all of these points in the 
future, strengthening the connection between consistent truncations and 
dualities even further.

\section*{Acknowledgements}
We would like to thank Riccardo Borsato, Sybille Driesen, Gabriel Larios, Gr\'egoire Josse, and Yuho Sakatani for helpful discussions. Parts of this work were finished while FH was visiting the group of Riccardo Borsato at the University of Santiago de Compostela. FH is very grateful for the hospitality he received during this time and for all the discussions from which this work benefited significantly. The work of FH is supported by the SONATA BIS grant 
2021/42/E/ST2/00304 from the National Science Centre (NCN), Polen. 
CNP is supported in part by DOE grant DE-FG02-13ER42020. 

\bibliography{literature}
   
\bibliographystyle{JHEP}

\end{document}